\documentclass[11pt,final]{siamonline0516}
\usepackage{amssymb}
\usepackage{amsmath}
\usepackage{amsfonts}
\usepackage{amscd}
\usepackage{graphicx}
\usepackage{epstopdf}
\usepackage{subfig}
\usepackage{algorithmic}

\newtheorem{remark}{Remark}
\numberwithin{equation}{section}
\numberwithin{theorem}{section}
\numberwithin{remark}{section}
\numberwithin{algorithm}{section}
\def\thefigure{\arabic{figure}}
\def\thetable{\arabic{table}}
\numberwithin{table}{section}
\numberwithin{table}{section}
\numberwithin{figure}{section}

\begin{document}

\title{Stochastic resin transfer molding process}
\author{M. Park\thanks{%
School of Mathematical Sciences, University of Nottingham, University Park,
Nottingham, NG7 2RD, UK} \and M.V.
Tretyakov\thanks{%
School of Mathematical Sciences, University of Nottingham, University Park,
Nottingham, NG7 2RD, UK, email: Michael.Tretyakov@nottingham.ac.uk}}
\maketitle

\begin{abstract}
We consider one-dimensional and two-dimensional models of the stochastic resin transfer molding process,
which are formulated as random moving boundary problems.
We study their properties, analytically in the one-dimensional case
and numerically in the two-dimensional case. We show how variability of time to fill depends on
correlation lengths and smoothness of a random permeability field.
\end{abstract}

\section{Introduction}

Over the last few decades the use of fiber-reinforced composite materials in
aerospace, automotive and marine industries, in sports and other areas
has seen a significant growth. The main benefits of composites include their
lightweight and high-performance nature, as well as their
flexibility to accommodate complex designs (see e.g. \cite{Ast97,AS10,Long05}
and references therein). One of the main manufacturing processes for
producing advanced composites is resin transfer molding (RTM), which belongs
to the Liquid Composite Molding class of composite manufacturing processes.
RTM has five main stages \cite{Ast97,AS10,LBA96,SIAMRev97}: (i)
manufacturing of a reinforcing preform (e.g., from carbon fiber, glass fiber
or other fabric); (ii) packing the preform in a closed mold, which has a cavity with
the shape of the designed part; (iii) injecting resin into the mold
cavity to fill empty spaces between fibers; (iv) resin curing (it can start
during or after the injection stage); and (v) demolding, i.e., taking the
solidified part, after completion of curing, from the mold. In this work we
will consider the third stage (filling the preform by resin) only and we
will, for simplicity, assume that resin curing starts after completion of
the filling, or in other words, that the filling process is much faster than
the curing, which is a common assumption for RTM.
The injecting resin stage is crucial for getting the expected properties of a material.

It is widely accepted (see e.g. \cite%
{EL06,EL062,Gommer13,MAL14,Matveev,PP99,SC09,XT04,Coloc11} and references therein) that
composite manufacturing processes are accompanied by uncertainties. The
origins of these uncertainties include (a) variability of fiber
placements due to imperfections of stages (i)-(ii) of RTM; (b)
variability of fiber properties; (c) variability of resin properties;
and (d) variability of environment during stages (i)-(iv) of RTM.
Deviations from the design caused by uncertainties can result in defects,
which have two manufacturing consequences \cite{AS10,MAL14}. First, to avoid
compromising performance of the material due to possible defects, one
usually uses more conservative designs, making composites thicker and hence
more expensive and less lightweight. Second, defects (e.g. dry spots) can
lead to a relatively large amount of scrap which increases the cost of the
material. Being able to quantify these uncertainties is highly important for further
advances of composite manufacturing.

The above-mentioned variabilities usually cause
uncertainty of permeability/hydraulic conductivity, which in its turn leads
to variability in mold filling patterns and filling times in RTM. In this
paper our main focus is on dependence of variability of filling times
 on properties of random fields which model uncertainty of
hydraulic conductivity.

Estimation of a filling time for each particular design of a composite part
is important for a successful manufacturing process. The filling time should
be neither too short nor too long. On the one hand, it should be
sufficiently long to allow adequate impregnation of the fibers. On the other
hand, a too long filling time can lead to premature gelation (or even
solidification) of the resin, which should be avoided as it is a source of
defects. Further, a large variability of the filling time is, practically,
highly undesirable as it affects robustness of the technological process,
making it difficult for automation and standardization (e.g., of an
operator's guidance). Thus, understanding of factors influencing filling
time variability is of importance for fiber-reinforced composite
manufacturing.

We start (Section~\ref{sec:1d}) with considering a stochastic
one-dimensional moving-boundary problem, where analytical analysis is
possible. The hydraulic conductivity is modelled as a stationary log-normal
random field. In particular, we observe that though the mean filling time
does not depend on correlation length of hydraulic conductivity, the filling
time variance (as a measure of variability) does depend on the correlation
length. We also consider dependence of variation of the filling time on
smoothness of the random hydraulic conductivity. In Section~\ref{sec:2d} we
formulate two variants of a two-dimensional moving-boundary problem with a random hydraulic
conductivity tensor. In some  previous related works (see e.g. \cite{TW01,Coloc11})
on stochastic moving-boundary problems, it was assumed that the
random hydraulic conductivity is isotropic. However, permeability (and hence
hydraulic conductivity) is anisotropic by design in most cases of practical interest
(see e.g.\cite{Dag,Rub03,PNT96}).
Moreover, geometric variability (i.e., deviations from the design such as variation of gaps
between fiber yarns, variation of width and angles of fiber yarns in
comparison with design, etc.) was observed in experiments (see, e.g. \cite%
{SS08, Gommer13, MAL14,Matveev} and references therein), which gives further evidence for the need to model
hydraulic conductivity as an anisotropic random field.
The main difficulty in modeling 2- and 3-dimensional resin transfer processes is the possibility of
dry spot appearances, i.e., forming enclosed areas with moving fronts behind the main front.
The enclosed areas contain air, which is a compressible medium, while resin can be considered incompressible.
A full description of this phenomenon requires a two-phase model involving both incompressible and compressible phases,
which is too computationally demanding. Here we work with a reduced model in
which the air entrapment is taken into account by modifying the boundary
conditions on the internal fronts. A discrete-type version of such a model was proposed in \cite{LBA96} (see also \cite{AS10}).
Here we formulate its continuous analogue. We note that though our study is motivated by technological processes in production of
composite materials, the considered models are also useful for other porous media problems (see \cite{Dag,Rub03,TW01} and references therein).
The two-dimensional model is solved numerically using the control volume-finite element method
(CV/FEM) which we present for completeness in Section~\ref{sec:alg}. The
results of our numerical study of the two-dimensional stochastic moving-boundary problem are given in
Section~\ref{sec:exp}, where we also experimentally examine  convergence of the CV/FEM algorithm
with the quantities of interest being the filling time and void content.
A discussion and concluding remarks are in Section~\ref{sec:end}.

\section{One-dimensional model\label{sec:1d}}

We start with studying a one-dimensional moving-boundary model for a
stochastic RTM process. In what follows we assume that we have a
sufficiently rich probability space $(\Omega ,\mathcal{F},P).$ Let $%
K(x)=K(x;\omega )$ be a random hydraulic conductivity defined on $[0,x_{\ast
}]\times \Omega $ and taking values on the positive real semi-line. Assuming
the resin's incompressibility, the moving boundary problem for the pressure of
resin $p(t,x)$ takes the form \cite{AS10,TW01}:
\begin{eqnarray}
-\frac{d}{dx}K(x)\frac{d}{dx}p(t,x) &=&0,\ 0<x<L(t),\ t>0,  \label{one1} \\
p(0,x) &=&p_{0},\ x\in (0,x_{\ast }],  \notag \\
p(t,0) &=&p_{I},\ t\geq 0,  \notag \\
\frac{d}{dt}L(t) &=&-\frac{K(L(t))}{\varkappa }\frac{d}{dx}p(t,L),\ \ L(0)=0,
\notag \\
p(t,L(t)) &=&p_{0},\ t>0,  \notag
\end{eqnarray}%
where $L(t)$ is the moving boundary, $\varkappa >0$ is the medium porosity, $%
p_{I}$ is a pressure on the inlet $x=0,$ and $p_{0}\leq p_{I}$ is the
pressure at the outlet $x=x_{\ast }$.

%\todo[inline]{In the Wikipedia page about permeability, the hydraulic conductivity expression also involves the dynamic viscosity of fluid. We need to think about this further.}
We recall that the hydraulic conductivity $K$ can be expressed as
\begin{equation}
K=\frac{k}{\mu },  \label{one101}
\end{equation}%
where $k$ is the permeability of the medium and $\mu $ is the viscosity of
the resin.

\begin{remark}
\label{rem:case}For simplicity we assume that the porosity $\varkappa $ in (%
\ref{one1}) is constant. Porosity can be assumed constant when its variability is significantly smaller than variability of hydraulic conductivity,
which is often the case (see e.g. \cite{Dag,Rub03}).
It is possible to modify the arguments of this
section for the case of random porosity but we do not consider it here.
Also, for definiteness, we impose the constant pressure condition at the
inlet but other boundary conditions (e.g. constant rate) can be also
considered.
\end{remark}

\begin{remark}
We note that without losing generality we can put $p_{0}=0$ in (\ref{one1})
(i.e., either it is vacuum on the outlet or $p$ is considered as the
relative pressure) but for the sake of the two-dimensional model considered
in the next section, it is convenient to keep the parameter $p_{0}.$
\end{remark}

Let
\begin{equation*}
F(y):=\int_{0}^{y}\frac{dz}{K(z)}.
\end{equation*}

\noindent \textbf{Assumption 2.1.} \textit{We assume that the random field }$%
K(x)=K(x,\omega ),$ $(x,\omega )\in \lbrack 0,x_{\ast }]\times \Omega ,$
\textit{is such that}

(i) $K(x)>0$ \textit{for} $x\in \lbrack 0,x_{\ast }]$ \textit{and all} $\omega \in \Omega ;$

(ii) \textit{the integral}
\begin{equation*}
G(x):=\int_{0}^{x}F(y)dy
\end{equation*}%
\textit{exists for} $x\in \lbrack 0,x_{\ast }]$ \textit{and a.e.} $\omega \in \Omega
. $ \medskip

It is not difficult to prove the following proposition (see similar
statements e.g. in \cite{AS10,TW01}).

\begin{proposition}
\label{prop1}Let Assumption~2.1 holds. Then the unique solution of (\ref%
{one1}) is
\begin{eqnarray}
L(t) &=&G^{-1}((p_{I}-p_{0})t/\varkappa ),  \label{one2} \\
p(t,x) &=&p_{I}-(p_{I}-p_{0})\frac{F(x)}{F(L(t))},\ t\geq 0,\ 0\leq x\leq
L(t),\ \ \ \text{a.s.}  \notag
\end{eqnarray}
\end{proposition}

Note that Assumption~2.1 implies that $G(x),$ $x\geq 0,$ is a.s. continuous
and strictly increasing, which guarantee existence of the inverse $%
G^{-1}(\cdot )$, needed for (\ref{one2}). We also remark in passing that if
dynamics of the front $L(t)$ are given then the nonlinear problem (\ref{one1}%
) becomes linear.

As we mentioned in the Introduction, from the application's point of view, an
important characteristic is the time $\tau =\tau (\omega )$ to fill a piece
of material of length $x_{\ast },$ i.e., the random time $\tau $ such
that
\begin{equation}
L(\tau )=x_{\ast }.  \label{one3}
\end{equation}%
It follows from (\ref{one2}) that
\begin{equation}
\tau =\frac{\varkappa G(x_{\ast })}{p_{I}-p_{0}}.  \label{one4}
\end{equation}

It is natural to assume that the mean of $K(x)$ corresponds to the hydraulic conductivity intended
by the design of a composite part and the perturbation of this mean is a stationary random field which models uncertainty due to the manufacturing process.
Let us now consider the case of the hydraulic conductivity $K(x)$ being a
stationary log-normal random field, which is a commonly used assumption
(see, e.g. \cite{PP99,Gommer13,MAL14,log12,Coloc11}), i.e.,
\begin{equation}
K(x)=K_{0}\exp (\varphi (x)),  \label{one5}
\end{equation}%
where $K_{0}>0$ and $\varphi (x)=\varphi (x;\omega ),$ $(x,\omega )\in
\lbrack 0,x_{\ast }]\times \Omega ,$ is a stationary Gaussian random field
with zero mean and covariance function $r(x).$
Note that
\begin{eqnarray}
EK(x) &=&K_{0}\exp \left( \frac{1}{2}r(0)\right) ,  \label{one51} \\
 \mathrm{Var} K(x) &=&K_{0}^{2}\exp \left( r(0)\right) \left[ \exp \left( r(0)\right) -1%
\right] ,\notag \\
\mathrm{Cov(}K(x),K(y)) &=&K_{0}^{2}\exp \left( r(0)\right) \left[ \exp
\left( r(x-y)\right) -1\right] .  \notag
\end{eqnarray}

The filling time according to the design is equal to
\[
\tau _{\text{designed}}=\frac{\varkappa x_{\ast }^{2}}{2(p_{I}-p_{0})EK}.
\]

The first condition in Assumption~2.1 is obviously satisfied by $K(x)$ from (%
\ref{one5}). To satisfy the second condition, it is sufficient to require
that realizations of $\varphi (x)$ are continuous with probability one and we make the
following assumption.\medskip

\noindent \textbf{Assumption 2.2.} \textit{We assume that the stationary
Gaussian random field }$\varphi (x)$ \textit{has zero mean and continuous
covariance function} $r(x)$ \textit{such that for some} $C>0$ \textit{and} $\alpha ,$ $\delta
>0:$%
\begin{equation}
r(0)-r(x)\leq \frac{C}{|\ln |x||^{1+\alpha }}  \label{one6}
\end{equation}%
\textit{for all }$x$ \textit{with} $|x|<\delta .$ \medskip

Under Assumption~2.2 the random field $\varphi (x)$ has continuous sample paths with
probability one (see e.g. \cite{Adler}).

For $K(x)$ from (\ref{one5}), we obtain the following statistical
characteristics of the filling time:
\begin{eqnarray}
E\tau &=&\frac{\varkappa }{2(p_{I}-p_{0})K_{0}}\exp \left( \frac{1}{2}%
r(0)\right) x_{\ast }^{2},  \label{one7} \\
\mathrm{Var}\tau &=&\frac{\varkappa ^{2}\exp \left( r(0)\right) }{%
(p_{I}-p_{0})^{2}K_{0}^{2}}\left[ \int_{0}^{x_{\ast }}\int_{0}^{x_{\ast
}}\int_{0}^{y}\int_{0}^{y^{\prime }}\exp \left( r(z-z^{\prime })\right)
dzdz^{\prime }dydy^{\prime }-\frac{x_{\ast }^{4}}{4}\right] .  \notag
\end{eqnarray}%
To understand the behavior of $E\tau $ and $\mathrm{Var}\tau $ in terms of $%
K(x),$ it is convenient to re-write (\ref{one7}) via the mean and variance of $K(x)$:
\begin{eqnarray}
E\tau &=&\frac{\varkappa x_{\ast }^{2}}{2(p_{I}-p_{0})}\frac{EK^{2}(x)}{%
\left[ EK(x)\right] ^{3}}=\frac{x_{\ast }^{2}\varkappa }{2(p_{I}-p_{0})EK(x)}%
\left[ \frac{\text{Var}K(x)}{\left[ EK(x)\right] ^{2}}+1\right] ,
\label{one70} \\
\mathrm{Var}\tau &=&\frac{\varkappa ^{2}\left[ \frac{\text{Var}K(x)}{\left[ EK(x)\right] ^{2}}+1\right]
^{2}}{(p_{I}-p_{0})^{2}\left[ EK(x)%
\right] ^{4}}\int_{0}^{x_{\ast }}\int_{0}^{x_{\ast }}\int_{0}^{y}\int_{0}^{y^{\prime
}}\mathrm{Cov(}K(z),K(z^{\prime }))dzdz^{\prime }dydy^{\prime }. \,\,\, \label{oneVar}
\end{eqnarray}%
It is interesting that the mean filling time depends only on Var$K(x)$ and $%
EK(x)$ and does not depend on the covariance, and hence it does not depend
on the correlation length and smoothness of the random field $K(x).$ We pay
attention to the interesting fact that $E\tau $ grows linearly with Var$%
K(x). $

The variance of the filling time $\mathrm{Var}\tau $ depends on covariance of $K(x)$.
Then to understand its behavior with respect to correlation length and
smoothness of $K(x),$ let us consider some particular cases of $\varphi (x).$

First let us look at the simple case of $\varphi (x)$ being independent of $%
x,$ i.e., being just a Gaussian random variable with zero mean and variance $%
\sigma ^{2}$ which can be viewed as a perfectly correlated medium. Then
\begin{equation}
\mathrm{Var}\tau =\frac{\varkappa ^{2}x_{\ast }^{4}}{(p_{I}-p_{0})^{2}\left[
EK(x)\right] ^{4}}\left[ \frac{\text{Var}K}{\left[ EK\right] ^{2}}+1\right]
^{2}\text{Var}K,  \label{one711}
\end{equation}%
i.e., $\mathrm{Var}\tau $ grows cubically with increase of Var$K$ in this
case.

\begin{remark}
In this case of a perfectly correlated medium we also have from (\ref{one4}):
\begin{eqnarray}
EK &=&\frac{\varkappa x_{\ast }^{2}}{2(p_{I}-p_{0})}E\frac{1}{\tau },
\label{one72} \\
\mathrm{Var}K &=&\frac{\varkappa ^{2}x_{\ast }^{4}}{4(p_{I}-p_{0})^{2}}%
\mathrm{Var}\frac{1}{\tau }.  \notag
\end{eqnarray}%
We note in passing that (\ref{one72}) is often used in experiments for
estimating an effective macroscopic hydraulic conductivity (and hence
effective permeability) via observing time to fill of samples of a material
(see e.g. \cite{Matveev}).
But it is not difficult to see that when the hydraulic
conductivity is not perfectly correlated, (\ref{one72}) might not be a good
way to estimate the mean and variance of hydraulic conductivity (cf. (\ref{one70})-(\ref{oneVar})).
\end{remark}

Now consider the following Mat\'{e}rn covariance function
for the stationary Gaussian random field $\varphi (x)$ (see \cite{matern1,yaglom1,Stein,matern2}):
\begin{equation} \label{eq:matern}
r(x) = \sigma^{2}\frac{1}{\Gamma(\nu) 2^{\nu-1}}\left( \sqrt{2\nu}d(x,\lambda)\right)^{\nu} K_\nu\left(\sqrt{2\nu}d(x,\lambda)\right),
\end{equation}
where $\Gamma(\cdot)$ is the gamma function, $K_\nu(\cdot)$ is the modified Bessel function of the second kind, $\sigma ^{2}$ is variance,
$\nu>0$ is a smoothness parameter, $\lambda >0$ is a characteristic
correlation length, and $d(x,\lambda)$ is a scaled distance function.
In particular, we have
\begin{eqnarray}
(\nu = 1/2) \qquad r(x) &=&\sigma ^{2}\exp \left( -d(x,\lambda)\right) ,  \label{ma1} \\
(\nu = 3/2) \qquad r(x) &=&\sigma ^{2}\left( 1+\sqrt{3}d(x,\lambda)\right) \exp \left( -%
\sqrt{3}d(x,\lambda)\right) ,  \label{ma2} \\
(\nu = 5/2) \qquad r(x) &=&\sigma ^{2}\left( 1+\sqrt{5}d(x,\lambda)+\frac{5}{3}d(x,\lambda)^2%
\right) \exp \left( -\sqrt{5}d(x,\lambda)\right)
,  \,\,\, \label{ma3} \\
(\nu \rightarrow \infty) \qquad r(x) &=&\sigma ^{2}\exp \left( -d(x,\lambda)^2\right) . \,\,\,
\label{mainf}
\end{eqnarray}%
The parameter $\nu$ in the above covariance functions controls the degree of smoothness of sample paths of the random field.
The random field $\varphi (x)$ with Mat\'{e}rn covariance function (\ref{eq:matern}) has $\lceil \nu -1 \rceil$ sample
path continuous derivatives with probability one.
Hence, with the exponential covariance function $r(x)$ from (\ref{ma1}), the random field $\varphi (x)$ has continuous
(but not differentiable) sample paths with probability one;
with $r(x)$ from (\ref{ma2}) --
once differentiable sample paths with probability one; with $r(x)$ from (\ref{ma3}) -- twice differentiable sample paths with probability one;
and with $r(x)$ from (\ref{mainf}) -- infinitely many times differentiable sample paths.

Note that we will use the four Mat\'{e}rn covariance functions in this section for $x$ being one-dimensional and in the next sections for
$x$ being two-dimensional. The scaled distance of the form
$d(x,\lambda) = |x|/\lambda$, with $|x|$ being the usual Euclidean
distance, corresponds to an isotropic random field.
In the two-dimensional case considered later in the paper we also model the random conductivity as an anisotropic
random field, appropriately choosing the scaled distance $d(x,\lambda)$ (see Section~\ref{sec:exp}).

\begin{figure}[th]
\centering
\includegraphics[width=0.6\textwidth]{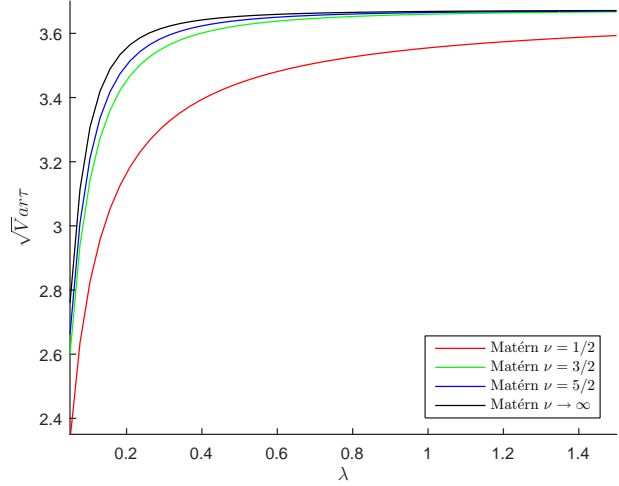}
\caption{ Dependence of the standard deviation $\sqrt{ \mathrm{Var}\tau }$ (in sec) of the filling time on the correlation length $\lambda$ (in m)
and on the smoothness parameter $\nu$. Here $x^* = 0.2$m, $\varkappa = 0.7$, $p_I-p_0 = 0.5$ MPa, $EK=10^{-8}$ m$^2$/sec$\cdot$Pa,
and $\sqrt{\mathrm{Var} K}= 8 \cdot 10^{-9}$ m$^2$/sec$\cdot$Pa (the corresponding $\sigma^2 \doteq 0.495$). The mean filling time $E\tau \approx 4.592$ sec and the filling time according to the design
$\tau _{\text{designed}} \approx 2.8 $ sec. }
\label{fig:vartau}
\end{figure}

We recall (cf. (\ref{one7})) that the mean filling time does not depend on the
choice of covariance and hence, in particular, it is the same for all four Mat\'{e}rn
covariance functions (\ref{ma1})-(\ref{mainf}). It is not difficult to show that,
in the case of these Mat\'{e}rn covariance functions, variance $\mathrm{Var}%
\tau $ of the filling time has the following properties: (i) it is
increasing with growth of the correlation length $\lambda $; (ii) for small
correlation lengths relative to the size of the material $x_{\ast }/\lambda
>>1,$ $\mathrm{Var}\tau $ grows linearly with $\lambda ;$ (iii) for large
correlation lengths $x_{\ast }/\lambda <<1,$ $\mathrm{Var}\tau $ is almost
independent of $\lambda $ (note that in the case of a perfectly correlated
medium we had (\ref{one711})); and (iv) for a fixed $\lambda ,$ variance $%
\mathrm{Var}\tau $ of the filling time grows with smoothness of the random
field. We illustrate these properties in Fig.~\ref{fig:vartau}.

As an example, we also give the expansion of $\mathrm{Var}\tau $  in the case of $\nu =1/2 $ and small $\sigma >0 $
(expansions in small $\sigma$ of statistical moments of the interface dynamics and of the pressure were derived in \cite{LPT05}):
\[
\mathrm{Var}\tau =\sigma ^{2}\frac{\varkappa ^{2}\exp \left( \sigma
^{2}\right) x_{\ast }^{4}}{(p_{I}-p_{0})^{2}K_{0}^{2}}\left[ \rho (\frac{%
\lambda }{x_{\ast }})+\frac{\sigma ^{2}}{2}\rho (\frac{\lambda }{2x_{\ast }})%
\right] +O(\sigma ^{6}),
\]%
where
\[
\rho (y)=\frac{2}{3}y-y^{2}-2y^{3}\exp (-1/y)-2y^{4}\exp (-1/y)+2y^{4}.
\]

To summarize, in the one-dimensional case,

\begin{itemize}
\item The mean filling time $E\tau $ does not depend on the correlation
length $\lambda $ or on smoothness of the random hydraulic conductivity $%
K(x)$. It decreases with increase of the mean $EK$ of the hydraulic
conductivity and with increase of $(p_{I}-p_{0})$; and it linearly increases
with increase of variance $\mathrm{Var}K$ of the hydraulic conductivity and
quadratically with increase of the length $x_{\ast }$.

\item The standard deviation of filling time $\sqrt{\mathrm{Var}\tau }$
grows as $\sqrt{\lambda }$ for small correlation lengths $x_{\ast }/\lambda
>>1$ and is almost independent of $\lambda $ for large correlation lengths $%
x_{\ast }/\lambda <<1$. For a fixed $\lambda ,$ it also grows with increase
of smoothness of the random field.
\end{itemize}

The important consequence of these
observations is significance of the correlation length for variability of
the filling time. Dependence of variability of $\tau $ on smoothness of $%
K(x) $ serves as a warning that stochastic modeling of permeability via homogenization
procedures needs to be done carefully. We also note that the mean filling time $E\tau$ is larger than
the filling time expected from the design. Further, standard deviation of the filling time is comparable
with the mean filling time, i.e. variability of the filling time is high, which can cause problems in
fiber-reinforced composite manufacturing as explained in the Introduction.

\section{Two-dimensional model\label{sec:2d}}

In this section we formulate the two-dimensional analog of the
one-dimensional model (\ref{one1}). To represent a mold, consider
an open two-dimensional domain
$D$ with the boundary $\partial D=\partial D_{I}\cup \partial D_{N}\cup
\partial D_{O}$, where $\partial D_{I}$ is the inlet, $\partial D_{N}$
is the perfectly sealed boundary, and $\partial D_{O}$ is the outlet.
Let $K(x,y)=K(x,y;\omega )$ be a random second-order hydraulic conductivity
tensor defined on $\bar{D}\times \Omega $. The moving-boundary problem for
the pressure of resin $p(t,x,y)$ takes the form (cf. \cite{AS10,TW01}):%
\begin{eqnarray}
-\nabla K(x,y)\nabla p &=&0,(x,y)\in D_{t},\ t>0,  \label{two1} \\
p(0,x,y) &=&p_{0},\ (x,y)\in D,  \notag \\
p(t,x,y) &=&p_{I},\ (x,y)\in \partial D_{I},\ t\geq 0,  \notag \\
\hat{n}(x,y)\cdot \nabla p(t,x,y) &=&0,\ (x,y)\in \partial D_{N},\ t\geq 0,
\notag \\
V(t,x,y) &=&-\frac{K(x,y)}{\varkappa }\hat{n}(t,x,y)\cdot \nabla p(t,x,y),\
(x,y)\in \Gamma (t),\ t\geq 0,  \notag \\
\Gamma (0) &=&\partial D_{I},  \notag \\
p(t,x,y) &=&p_{0},\ (x,y)\in \Gamma (t),\ t>0,  \notag\\
p(t,x,y) &=&p_{0},\ (x,y)\in \partial D_{O},\ t\geq 0,  \notag
\end{eqnarray}%
where $V(t,x,y)$ is the velocity of the moving boundary $\Gamma (t)$ in the
normal direction and $\hat{n}(x,y)$ and $\hat{n}(t,x,y)$ are the unit outward
normals to the corresponding boundaries,
$D_t \in D$ is the time-dependent domain bounded by the moving boundary $\Gamma(t)$
and the appropriate parts of $\partial D$,
$\varkappa >0$ is the medium
porosity, $p_{I}$ is a pressure at the inlet $\partial D_{I},$ and $%
p_{0}\leq p_{I}$ is the pressure at the outlet $\partial D_{O}$. Remark~\ref%
{rem:case} is applicable here.

\begin{figure}[th]
\centering
\includegraphics[width=0.6\textwidth]{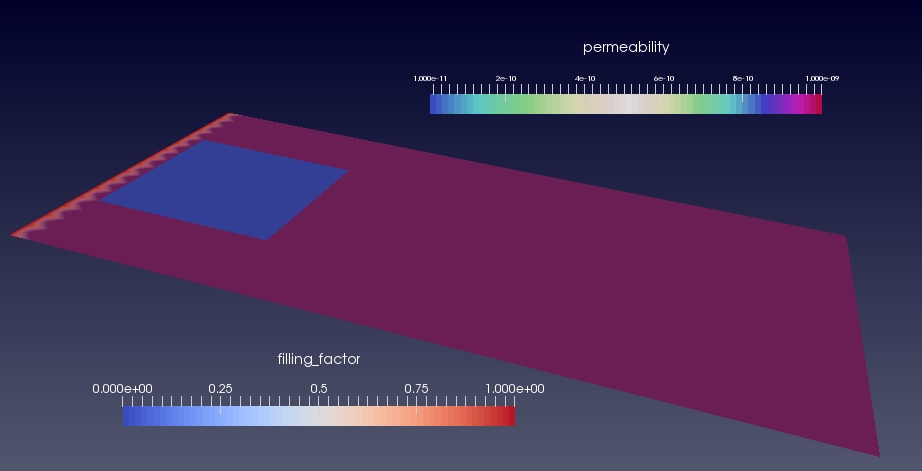}
\caption{The 2m by 1m rectangular domain $\bar D$ with inlet $\partial D_{I}$
being the left side of $\bar D$ and outlet $\partial D_{O}$ being the right side
of $\bar D$. The top and the bottom sides of $\bar D$ form the perfectly sealed
boundary $\partial D_{N}$. The small domain with low hydraulic conductivity has the
size 0.5m by 0.5m. Here $K_1=10^{-7}$ m$^2$/sec$\cdot$Pa
and $K_2=10^{-9}$ m$^2$/sec$\cdot$Pa, see further details in the text.}
\label{fig:void1}
\end{figure}

\begin{figure}[th]
\centering
\includegraphics[width=0.45\textwidth]{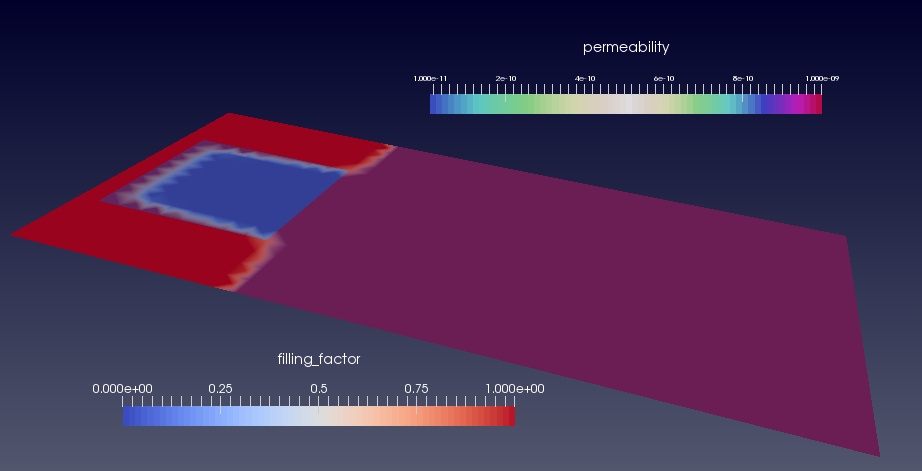} %
\includegraphics[width=0.45\textwidth]{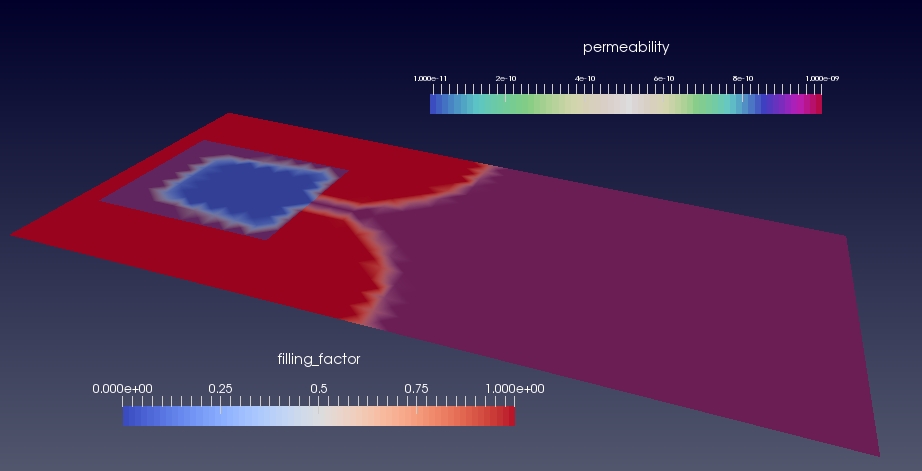} %
\includegraphics[width=0.45\textwidth]{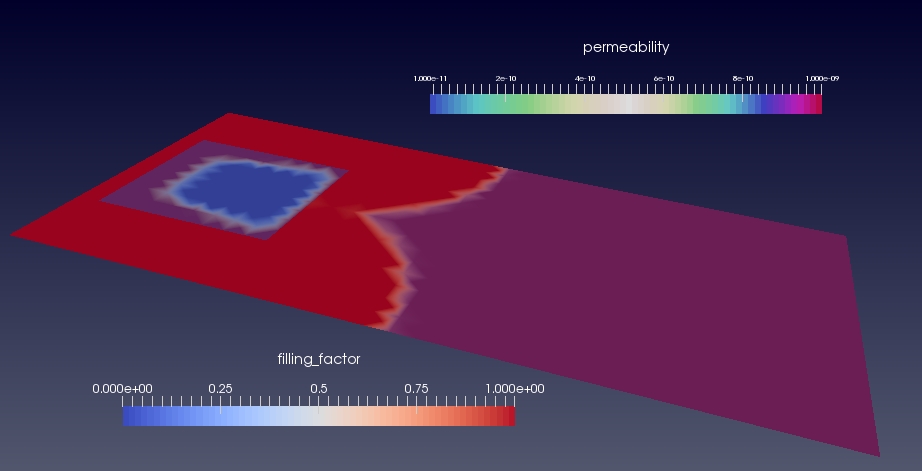}
\caption{Void formation. The thickness of the material is 1cm, the inlet
pressure $p_I=0.6 MPa$, the outlet pressure $p_0=0.1 MPa$, and the porosity $%
\varkappa=0.7$. The
other parameters are given in Fig.~\protect\ref{fig:void1}.}
\label{fig:void2}
\end{figure}

Behavior of the two-dimensional model (\ref{two1}) is considerably more
complex than of the one-dimensional model (\ref{one1}). Let us start with an
illustrative example (see also e.g. \cite[Ch. 8]{AS10}). Consider a
rectangular piece of material which has hydraulic conductivity $K(x,y)=K_{1}I$
constant everywhere in the domain $\bar D$ (here $I$ is the $2\times 2$ unit
matrix) except a relatively small region $D_{low}\ $ which has (again constant)
hydraulic conductivity $K(x,y)=K_{2}I$, where $K_{2}\ll K_{1}$ (see Fig.~\ref%
{fig:void1}). In this case the resin can race around the low permeability
region and the front becomes discontinuous, creating a macroscopic void
behind the main front as demonstrated in Fig.~\ref{fig:void2}. Based on the
one-dimensional model (\ref{one1}) and Proposition~\ref{prop1}, it is not
difficult to estimate that it is sufficient to have $K_{2}/K_{1} < 1/9$
for appearance of a void. Macroscopic voids are one of the main defects in
composites leading to scrap and failures (see e.g. \cite{AS10} and
references therein). Possible discontinuities of the front cause
difficulties in both analytical analysis of (\ref{two1}) and its numerical
approximation as we discuss further in Sections~\ref{sec:alg} and~\ref{sec:conv}.

In practice one does not have a deep vacuum (i.e., $p_{0}$ cannot be assumed
negligible) and air is entrapped in macrovoids \cite{AS10}. To take into
account air entrapment, one can replace the model (\ref{two1}) by a
two-phase model with one phase (resin) being incompressible (as it is in (%
\ref{two1})) and the other (air) being compressible. But such a model is
computationally expensive, while to find an optimal design for a composite
part's production (e.g. optimal locations of vents and inlets), especially
taking into account uncertainties, one needs to run model simulations very
many times. To this end a simplified model is considered \cite{LBA96,AS10},
in which the air entrapment is taken into account by modifying the boundary
conditions on the internal fronts. Such a model is widely used in the
engineering community working on advanced composite manufacturing, including
related commercial software (see, e.g. \cite{AS10} and references therein).
In the simplified model it is assumed \cite{LBA96,AS10} that the pressure in
a void increases according to the ideal gas law, i.e., that pressure inside
the void multiplied by its volume remains constant when the void shrinks
during the filling process (we assume that temperature remains constant
during the process). In \cite{LBA96,AS10} there is a discrete-type formulation
of this model, here we give its PDE formulation.

To describe this modification of (\ref{two1}), assume that at time $t\geq 0$
there are $\ell (t)$ entrapments with closed boundaries $\Gamma _{i}(t)$ and
volumes $v_{i}(t)$ formed at times $\tau _{i},$ $i=1,\ldots ,\ell (t),$
behind the main front $\Gamma _{0}(t).$ Then we can write the model as
\begin{eqnarray}
-\nabla K(x,y)\nabla p &=&0,(x,y)\in D_{t},\ t>0,  \label{two3} \\
p(0,x,y) &=&p_{0},\ (x,y)\in D,  \notag \\
p(t,x,y) &=&p_{I},\ (x,y)\in \partial D_{I},\ t\geq 0,  \notag \\
\hat{n}(x,y)\cdot \nabla p(t,x,y) &=&0,\ (x,y)\in \partial D_{N},\ t\geq 0,
\notag \\
V_{i}(t,x,y) &=&-\frac{K(x,y)}{\varkappa }\hat{n}_{i}(t,x,y)\cdot \nabla
p(t,x,y),\ (x,y)\in \Gamma _{i}(t), \notag \\
& & \qquad \qquad \hspace{1.1in} i=0,\ldots ,\ell (t),\ t\geq 0,  \notag
\\
\Gamma _{0}(0) &=&\partial D_{I},  \notag \\
p(t,x,y) &=&p_{0},\ (x,y)\in \Gamma _{0}(t),\ t>0,  \notag \\
p(t,x,y) &=&p_{0},\ (x,y)\in \partial D_{O},\ t\geq 0,  \notag \\
p(t,x,y) &=&\frac{p_{0}v_{i}(\tau _{i})}{v_{i}(t)},\ (x,y)\in \Gamma
_{i}(t),\ t>0,  \notag
\end{eqnarray}%
where $V_{i}(t,x,y)$ are the velocities of the moving boundaries
$\Gamma_{i}(t)$ in the normal direction and $\hat{n}(x,y)$ and $\hat{n}_{i}(t,x,y)$
are the unit normals to the corresponding boundaries,
$D_t \in D$ is the time-dependent domain bounded by $\Gamma_i(t)$, $i=0,\ldots ,\ell (t)$,
and the appropriate parts of $\partial D$,
the rest of the
notation is as in (\ref{two1}). The volume $v_{i}(t)$ of a void is computed as
\[
v_{i}(t)=H \int_{\Gamma_i(t)} d\Gamma_i(t),
\]
where $H$ is a fixed thickness of the material, which is assumed to be small so that flow through thickness
can be neglected, i.e., that the two-dimensional model is a good approximation for the three dimensional flow.

It is clear that in the model (\ref{two3}) once a void is formed at $\tau _{i}$, its volume $v_{i}(t)$ remains larger than or equal
to $p_{0}v_{i}(\tau _{i})/p_{I}$ and, consequently, the number of voids $\ell (t)$ is a non-decreasing function. Hence we have the following
inequalities for a void's volume:
\begin{equation}
\frac{p_0}{p_I} v_{i}(\tau_i) \leq v_{i}(t) \leq v_{i}(\tau_i), \, t \geq \tau_i,
\label{ieq1}
\end{equation}
and for the void content at a fixed time $T$:
\[
\frac{p_0}{p_I} \sum_{i=0}^{\ell (T)} v_{i}(\tau_i) \leq \sum_{i=0}^{\ell (T)} v_{i}(T)
\leq \sum_{i=0}^{\ell (T)} v_{i}(\tau_i).
\]

In the case of constant hydraulic conductivity, $K(x)=K$, (such a problem often called the Hele-Shaw problem or the
quasi-stationary Stefan problem) existence and uniqueness of (\ref{two1}) have been established
both locally and globally and in the classical and weak senses, see \cite{exi00,exi1,exi2,exi3,exi0,exi4} and also references therein.
Note that in this case no voids can be formed and (\ref{two1}) and (\ref{two3}) coincide.
The models (\ref{two1}) and (\ref{two3}) with non-constant $K(x)$ can have singular-type
behavior when voids form and, in the case of (\ref{two1}), also when they collapse.
Since there is no collapse of voids in (\ref{two3}), behavior of its solutions is less singular than for (\ref{two1}) and
in this sense (\ref{two3}) can be viewed as a regularization of (\ref{two1}).
We are not aware that questions concerning existence and uniqueness of solutions to (\ref{two1}) and (\ref{two3}) have been
considered in the literature, and they are an interesting and important topic
for further study, especially taking into account wide use of such models in applicable sciences.
Local existence and uniqueness of solutions to (\ref{two1}) and (\ref{two3}) can potentially be
addressed under some regularity of the data similarly to \cite{Rad92,Rad93}.

In most cases of practical interest permeability $k(x,y)$ (and hence the
hydraulic conductivity $K(x,y)$) is anisotropic \cite{Dag,PNT96,SS08, Gommer13, MAL14,Matveev}.
Consequently, it is important to model the hydraulic conductivity $K(x,y)$ as a random tensor.
The principal-axis transformation of the hydraulic conductivity tensor gives
\begin{equation}
K(x,y)=\mathbb{T}(x,y)\left[
\begin{array}{cc}
K_{xx}(x,y) & 0 \\
0 & K_{yy}(x,y)%
\end{array}%
\right] \mathbb{T}^{\intercal }(x,y),  \label{two4}
\end{equation}%
where $\mathbb{T}(x,y)$ is the rotation matrix%
\begin{equation*}
\mathbb{T}(x,y)=\left[
\begin{array}{cc}
\cos \theta (x,y) & \sin \theta (x,y) \\
-\sin \theta (x,y) & \cos \theta (x,y)%
\end{array}%
\right] .
\end{equation*}

We assume that $K_{xx}(x,y)=K_{xx}(x,y;\omega )$ and $%
K_{yy}(x,y)=K_{yy}(x,y;\omega )$ can be modeled as independent log-normal
random fields and the angle $\theta (x,y)=\theta (x,y;\omega )$ is a
Gaussian field independent of $K_{xx}(x,y)$ and $K_{yy}(x,y).$ We propose
that the means of $K_{xx}(x,y),$ $K_{yy}(x,y)$ and $\theta (x,y)$ correspond
to the hydraulic conductivity intended by the design of a composite part and
the perturbation of these means are stationary random fields modeling
uncertainty arising during the manufacturing process. Properties of the
model (\ref{two3}), (\ref{two4}) are discussed in Section~\ref{sec:exp}
below based on its simulation by the CV/FEM algorithm described in the next
section.

\section{Numerical algorithm\label{sec:alg}}
In this section, for completeness of exposition, we give an implementation of the interface-tracking control volume finite element method (CV/FEM) with a fixed grid \cite{AS10,Hirt81,cvfem1}, in a form suitable for the considered stochastic model (\ref{two3}), (\ref{two4}). CV/FEM is a volume-of-fluids technique \cite{Hirt81,vof}. It is widely used in the simulation of the RTM filling process \cite{AS10,LBA96,Bruschke:1990ii,Phelan:1997hy}. There are a number of alternatives to CV/FEM, including level set methods \cite{levelset}, other volume-of-fluids methods \cite{vof}, marker particle methods (see \cite{levelset} and references therein), and boundary element methods (see e.g. \cite{zabaras} and references therein).
Fixed-grid CV/FEM is currently the method of choice in the RTM community due to its computational efficiency \cite{AS10} and we follow this common RTM practice here. At the same time, it is of interest in future work to compare CV/FEM with modern level set methods.

Let us turn to the CV/FEM description. The whole computational domain (an empty mold) is first discretized using triangular elements, and then each element is further divided into three sub-volumes by connecting the center point and the midpoints of the edges of triangle. Each node is surrounded by a control volume that is composed of all of the sub-volumes associated with that node. Note that the number of control volumes is equal to the total number of nodes.

\begin{figure}
\centering
\subfloat[$CV_i$ contains two subelements from the bottom layer and one from the top layer.]{\includegraphics[width = .4\textwidth]{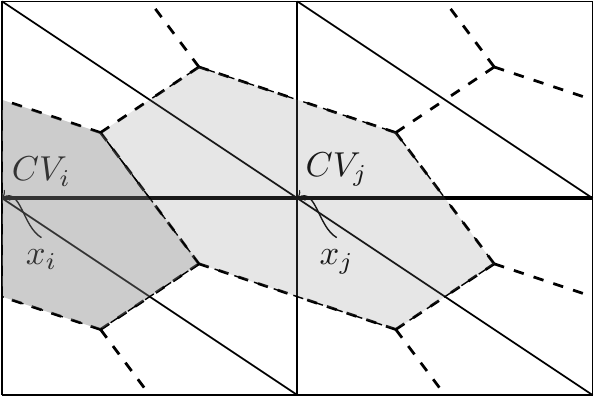}}
\hspace{1cm}
\subfloat[$CV_i$ contains two subelements from the bottom layer and another two from the top layer.]{\includegraphics[width = .4\textwidth]{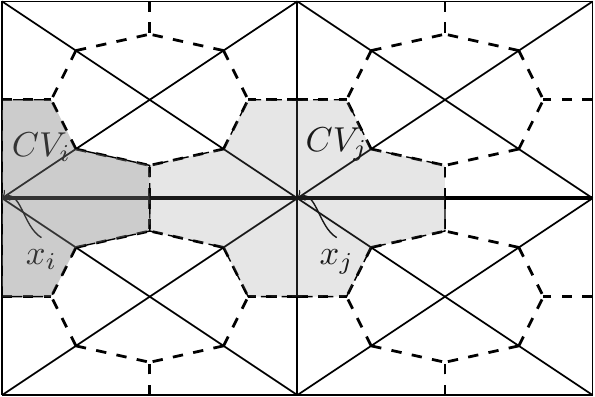}}
\caption{Two examples of tessellation of the spatial domain and the corresponding control volume subdivisions.
} \label{fig:control_volume}
\end{figure}

Figure \ref{fig:control_volume} shows two different types of tessellation of the spatial domain and the corresponding control volume subdivisions.  Suppose that the two horizontal layers divided by a thick line in the middle domain as in Fig.~\ref{fig:control_volume} have two different permeability values. It is not difficult to see (and was checked experimentally) that in the case of the discretization as in Fig.~\ref{fig:control_volume}(a) the exchange of permeability values between the two layers can result in different filling times due to the discretization asymmetry: the control volume $CV_i$, which has a boundary edge, has two subelements from the bottom layer and one from the top layer. In order to avoid this bias, we choose  unbiased (i.e. symmetric) control volumes as shown in Fig.~\ref{fig:control_volume}(b) for our numerical experiments.

In the CV/FEM, to track the interface, a scalar parameter, $f_i$, called the \textit{fill factor}, is assigned to each control volume $CV_i$. The fill factor represents the ratio of the volume of fluid to the total volume of the control volume. The fill factor $f_i$ takes values from 0 to 1: $f_i = 1$ for saturated region, $f_i = 0$ for an empty region and $0<f_i<1$ for a partially filled region. If $0 \leq f_i<1$, we will say that the control volume is unsaturated.
The flow front can be reconstructed based on the nodes that have partially filled control volumes, i.e., those with $0<f_i<1$
(see e.g. \cite{Hirt81,AP91}). In this paper we are not interested in reconstruction of the front and hence we do not consider it here. Finding all parts of the unsaturated regions requires a void detection algorithm. Such an algorithm was introduced in \cite{LBA96} and here we present its implementation. For simplicity, we assume that there is a single vent in the mold but it is not difficult to generalize the algorithm to the case of many vents.

\begin{figure}[h!]
\centering
\subfloat[The fill factors of 13 control volumes. Note that $CV_3, CV_6$, and $CV_9$ are the control volumes containing nodes on the vent.]{\includegraphics[width = .5\textwidth]{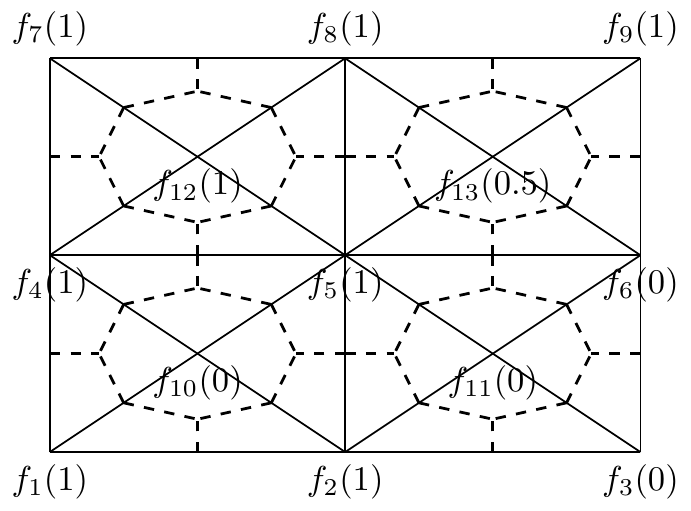}}
\\
\subfloat[Add vent control volumes $CV_3$ and $CV_6$ into the array from$\_$vent, as $f_3<1$ and $f_6<1$. ]{\includegraphics[width = .4\textwidth]{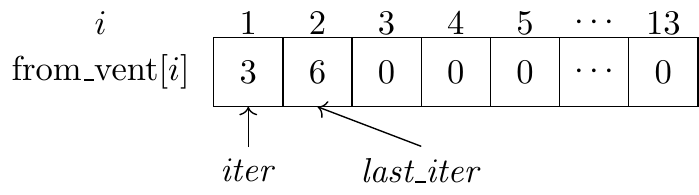}}
\hspace{1cm}
\subfloat[Add an unsaturated neighbour of $CV_{3}$, which is $CV_{11}$ in this case, into the array from$\_$vent and advance \textit{last$\_$iter} to the next element.] {\includegraphics[width = .4\textwidth]{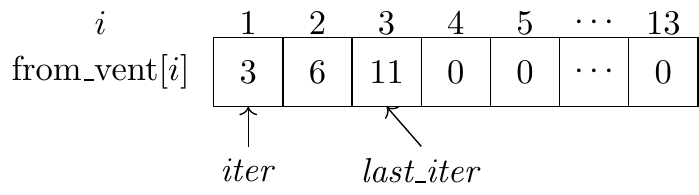}}
\\
\subfloat[First advance \textit{iter} to the next element. Then add an unsaturated neighbour of $CV_{6}$ into the array from$\_$vent and advance the \textit{last$\_$iter} to the next element.] {\includegraphics[width = .4\textwidth]{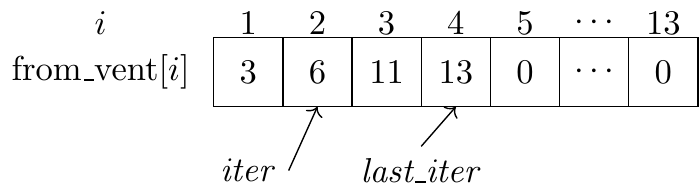}}
\hspace{1cm}
\subfloat[Advance \textit{iter} to the next element. There is no unsaturated neighbour of $CV_{11}$.] {\includegraphics[width = .4\textwidth]{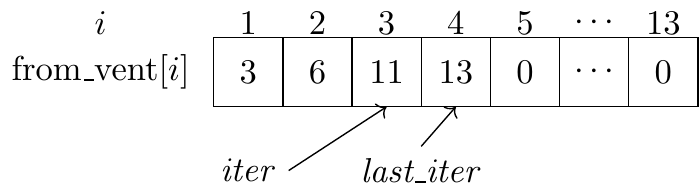}}
\\
\subfloat[Advance \textit{iter} to the next element. There is no unsaturated neighbour of $CV_{13}$.] {\includegraphics[width = .4\textwidth]{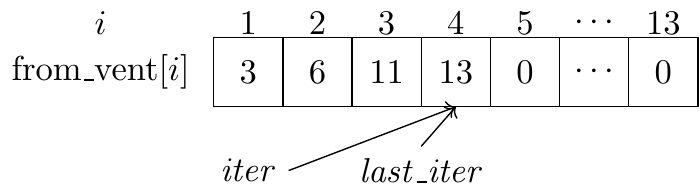}}
\hspace{1cm}
\subfloat[Advance  \textit{iter} to the next element. Now \textit{iter} $>$ \textit{last}$\_$\textit{iter}.] {\includegraphics[width = .4\textwidth]{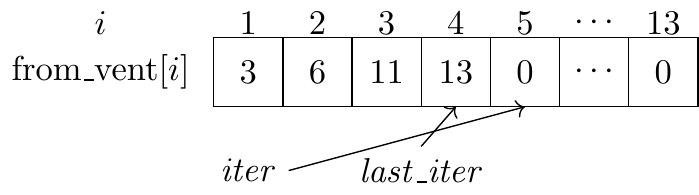}}
\caption{An example of a process to find all unsaturated control volumes connected to the vent through unfilled control volumes. } \label{fig:void_detection}
\end{figure}

 By a void control volume, we understand an unsaturated control volume which is not connected to a vent. To determine which control volumes belong to voids, we first find all unsaturated control volumes which are connected to the vent (i.e., which are not voids)
 through unsaturated control volumes. To do this, we introduce an array of  size equal to the number of nodes, \textit{from$\_$vent} and two pointers, \textit{iter} and \textit{last$\_$iter}, pointing at the first element and the last nonzero element of the array, respectively. This process is done in four steps (see Algorithm~\ref{alg:void_cv}), and Figure~\ref{fig:void_detection} illustrates it with an example.

\begin{algorithm}
\caption{Detection of control volumes which are connected to the vent}
	\label{alg:void_cv}
\begin{itemize}
\item[\textit{Step} 1] Add all unsaturated control volumes where the vent is located into the array \textit{from$\_$vent,} and set two pointers \textit{iter} and \textit{last$\_$iter} (Figure \ref{fig:void_detection} (b)).
\item [\textit{Step} 2] The loop is carried out over the neighbors of a control volume $CV_i$ pointed at by the pointer $iter$ to detect first visited neighbors whose fill factors are less than 1. Each time the neighbor is chosen to be added to \textit{from$\_$vent}, move \textit{last$\_$iter} to the next element (Figure \ref{fig:void_detection} (c)).
\item [\textit{Step} 3] After all neighbors of $CV_i$ are checked, advance \textit{iter} to the next element (Figure \ref{fig:void_detection} (d)).
\item [\textit{Step} 4] Repeat \textit{Step} 2 and \textit{Step} 3 until \textit{iter} $>$ \textit{last$\_$iter} (Figure \ref{fig:void_detection} (g)).

\end{itemize}

\end{algorithm}

Now the void control volumes are all unsaturated control volumes that do not belong to the \textit{from$\_$vent} formed by Algorithm~\ref{alg:void_cv}.
If there are void control volumes at the current time step $t_k$, then each individual void is represented by a connected set of void control volumes. Starting from any of the void control volumes, $CV_i$, a search process (analogous to Algorithm~\ref{alg:void_cv}) is used to find all void control volumes connected to $CV_i$ through void control volumes. The total volume of each void $j$ is computed during the search. If the void is formed for the first time at the time $t_k$ then we store its volume in $\bar v_j^*$; otherwise, assign it to  $\bar v_j(t_k)$.  Then the pressure value, $\bar p_{\mathrm{void}_j}(t_k)$,
in each void (including its boundary) at the time $t_k$ is determined using the ideal gas law as in (\ref{two3}):
\begin{equation}\label{eq:idealgas}
\bar p_{\mathrm{void}_j}(t_k) = \frac{p_0 \bar v_j^*}{\bar v_j(t_k)}.
\end{equation}

 \begin{figure}
\centering
\includegraphics[width = .5\textwidth]{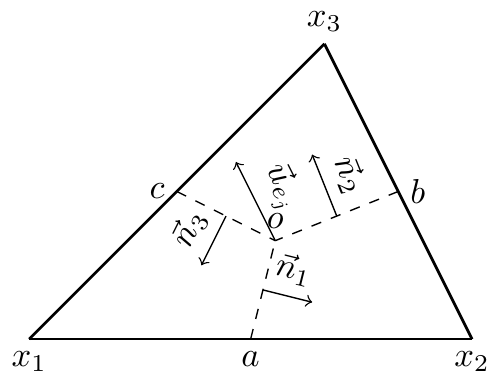}
\caption{Calculation of the local flow rate on a triangular element $e_j$ with nodes $x_1, x_2, x_3$. } \label{fig:flow_rate}
\end{figure}

At each time step $t_k$ of the CV/FEM, the fully-saturated control volumes form the solution domain and the finite element method is used to approximate the pressure field $\bar p$ in the solution domain. The corresponding boundary conditions on the inlet $\partial D_{I}$ and the perfectly sealed boundary $\partial D_{N}$ are imposed. Further, the pressure on the outlet $\partial D_{O}$ and in all-partially filled or empty control volumes not belonging to voids
 is set to $p_0$ while the pressure inside voids (i.e., for all unsaturated control volumes belonging to the voids) is set to $p_0$ if the void is formed at this time step and to $\bar p_{\mathrm{void}_j}(t_k)$ otherwise.

After computing the approximate pressure field $\bar p$, we calculate the velocity field for completely filled control volumes adjacent to unsaturated control volumes at the centroid of each element using the Darcy's law:
\begin{equation}\label{eq:darcy}
u(t,x,y) = -\frac{K(x,y)\cdot \nabla \bar p(t,x,y)}{\varkappa },
\end{equation}
where $u$ is the superficial fluid velocity.

Assuming that the fluid velocity is constant throughout each element, we compute the local flux into each subelement. Figure~\ref{fig:flow_rate} illustrates the local flux calculation in the triangular element $e_i$. The midpoints of the three sides of $e_i$ are denoted as $a$, $b$, and $c$ and point $o$ is the center of the element. The vectors $\vec{n}_1$, $\vec{n}_2$, and $\vec{n}_3$ are unit normal vectors perpendicular to the surfaces $oa$, $ob$, and $oc$, respectively.   The local flux into each subelement associated with a node $x_i$ in element $e_j$, $Q_{e_j,x_i}$, $i=1,2,3$, is calculated as
\begin{gather}
 Q_{e_j,x_1}  =  -u_{e_j}\cdot (\vec{n}_1|oa| - \vec{n}_3|oc|)H, \,\,\,\, Q_{e_j,x_2}  =  -u_{e_j}\cdot (\vec{n}_2|ob| - \vec{n}_1|oa|)H,  \label{eq:local_q} \\
 Q_{e_j,x_3}  = -u_{e_j}\cdot (\vec{n}_3|oc| - \vec{n}_2|ob|)H, \notag
\end{gather}
where H is the thickness of the mold. The total fluxes entering into the control volume $CV_i$ is then calculated by assembling the local fluxes:
\begin{equation}\label{eq:global_q}
Q_i = \sum_{e_l \in E_i}Q_{e_l,x_i},
\end{equation}
where $E_i$ is the set of elements containing the node $x_i$.

Having computed all the fluxes, we calculate the new fill factor $f_i(t_{k+1})=f_i(t_k+\delta t)$ of the unsaturated control volume $i$ as
\begin{equation}\label{eq:fill_factor}
f_i(t_k+\delta t_k) = f_i(t) + \frac{\delta t_k Q_i(t)}{\mathcal{V}_i},
\end{equation}
where $\mathcal{V}_i$ is the volume of control volume $i$. The time increment $\delta t_k$ is calculated so that at least one control volume is filled during the current time step and
\begin{equation}\label{eq:deltat}
\delta t_k = \min_i \frac{(1-f_i(t_k))\mathcal{V}_i}{Q_i},
\end{equation}
where the minimum is taken over all partially-filled control volumes and their neighboring empty control volumes (i.e., all the control volumes which are in a neighborhood of the moving front) at time $t_k$. Note that the element on which the minimum is reached becomes filled at $t_{k+1}=t_k+\delta t_k$.

This simulation process continues for every time step until one of two conditions is met: either (1) the entire mold is filled or (2) there is an equilibrium of pressure between the interior and the exterior of all voids and the rest of the mold is filled.
The corresponding time $\tau_h$ is considered as the approximate filling time obtained with the mesh size $h$ ($h$
is the length of hypotenuse of the triangular element).
To summarize, the CV/FEM is presented in Algorithm~\ref{alg:cvfem}.
\begin{algorithm}
\caption{Control volume finite element method (CV/FEM)}
	\label{alg:cvfem}
	\begin{algorithmic}
	 \STATE\textit{Step} 1. Create control volumes. Set $t_0=0$ and $k=0$.
	 \STATE\textit{Step} 2. Identify the saturated domain by the fill factors and detect void control volumes using Algorithm~\ref{alg:void_cv}.	
	 \STATE\textit{Step} 3. Compute the pressure field $\bar p$ in the saturated domain with the appropriate boundary conditions by the FEM.
	 \STATE\textit{Step} 4. Calculate the pressure gradients, $\nabla \bar p$, in the neighborhood of the flow front
                           by differentiating the element shape functions and then compute the velocity field, $u$, using Darcy's law (\ref{eq:darcy}).
	 \STATE\textit{Step} 5. Calculate the volumetric flow rate, $Q$, as in (\ref{eq:local_q}) and (\ref{eq:global_q}).
	 \STATE\textit{Step} 6. Find the size of the time step $\delta t_k$ by (\ref{eq:deltat}).
	 \STATE\textit{Step} 7. Update the fill factors with (\ref{eq:fill_factor}) and advance the time: $t_{k+1} = t_k + \delta t_k$ and set $k=k+1$.
	 \STATE\textit{Step} 8. Repeat from \textit{Step} 2 for the newly-filled domain until the mold is completely saturated
or there is an equilibrium of pressure between the interior and the exterior of all voids and the rest of the mold is filled.
	\end{algorithmic}
\end{algorithm}

To simulate the stochastic model (\ref{two3}), (\ref{two4}), we need to generate the random hydraulic conductivity tensor $K(x,y)$ at the centers of triangular elements, which requires to sample from stationary Gaussian distributions according to the proposed stochastic model for $K(x,y)$
at the end of Section~\ref{sec:2d}. To generate the required Gaussian random fields, we exploit the block circulant embedding method from \cite{PARK:2015kg}, which is an extension of the classical circulant embedding method \cite{Dietrich:1997ep,Wood:2012be} from regular grids to block-regular grids.

\section{Numerical experiments\label{sec:exp}}

In this section, we present results of numerical experiments which aim at (i) examining  convergence of Algorithm~\ref{alg:cvfem} for the model (\ref{two3}) with the quantities of interest being the filling time and void content (Section~\ref{sec:conv}); and (ii) studying how variability  of permeability affects the filling time in the RTM process (Sections~\ref{sec:num1d}-\ref{sec:num2d2}). To experimentally study convergence in Section~\ref{sec:conv}, we use the deterministic version of the model (\ref{two3}), i.e., when the hydraulic conductivity $K(x,y)$ is a given deterministic tensor. In the stochastic experiments in Sections~\ref{sec:num1d}-\ref{sec:num2d2} we consider the model (\ref{two3}), (\ref{two4}) with various parameters of the random hydraulic conductivity $K(x,y)$. The principal conductivity values,  $K_{xx}$ and $ K_{yy}$, and the angle $\theta(x,y)$  are assumed to be independent stationary random fields, with $K_{xx}$ and $ K_{yy}$ being log-normal and $\theta(x,y)$ being Gaussian. In our experiments the three Mat\'{e}rn  covariance functions (\ref{ma1})-(\ref{ma3}) for the Gaussian random fields $\log K_\alpha$,  $\alpha = xx, yy$, and $\theta$, are used with the following scaled distance
\begin{equation}
d(\textbf{x},\lambda) = \sqrt{\left(\frac{\textbf{x}_x^2}{\lambda_x}\right)^2+\left(\frac{\textbf{x}_y^2}{\lambda_y}\right)^2},
\, \textbf{x}=(\textbf{x}_x,\textbf{x}_y).
\end{equation}
In Sections~\ref{sec:num1d}-\ref{sec:num2d2} we will denote the means of $K_{xx}$, $ K_{yy}$ and $\theta$ by $\mu_{K_{xx}}$, $\mu_{K_{yy}}$ and $\mu_{\theta}$, respectively, and the standard deviations of $K_{xx}$, $ K_{yy}$ and $\theta$ by $\sigma_{K_{xx}}$, $\sigma_{K_{yy}}$ and $\sigma_{\theta}$, respectively.  In each particular experiment the covariance function and the correlation lengths $\lambda_x$ and $\lambda_y$ will be chosen to be the same for all three random fields.
To compute expectation and variance of the filling time  $\tau$, we exploit the Monte Carlo technique using $3000$ independent runs of (\ref{two3}), (\ref{two4}) in all the Monte Carlo experiments presented in Sections~\ref{sec:num1d}-\ref{sec:num2d2}.

\begin{figure}
\centering
\includegraphics[width = .9\textwidth]{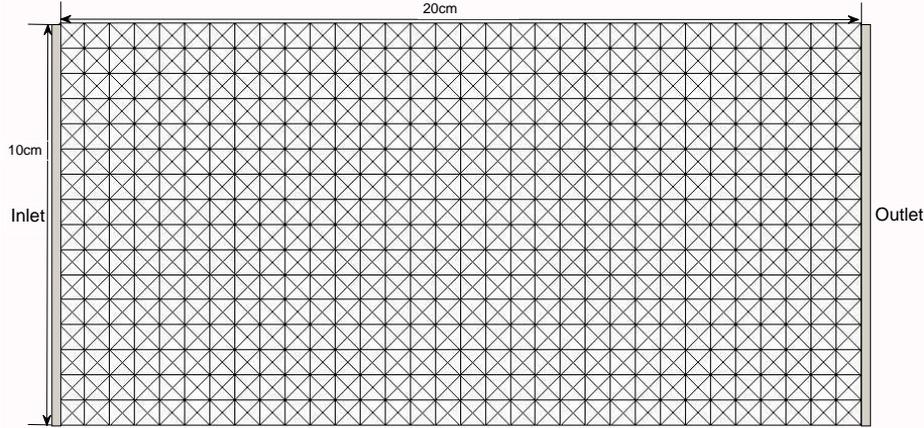}
\caption{Schematic illustration of the mold geometry with an inlet gate and an outlet gate and tessellation of the spatial domain.} \label{fig:sim_grid}
\end{figure}

In all our experiments the resin is injected into a rectangular mold of size 20cm$\times$10cm$\allowbreak\times$0.1cm as shown in Figure \ref{fig:sim_grid}. The horizontal direction is viewed as the $x$-direction and vertical as the $y$-direction.
Since in this work we assume that there is no flow in the thickness direction, we use two-dimensional elements for modeling the flow. In what follows we denote the mesh size by $h$, which is the length of hypotenuse of the triangular element. The physical properties of the preform are chosen in all the experiments, except the ones with a void in Section~\ref{sec:conv}, as listed in Table~\ref{tab:properties}.

\begin{table}[h!]
\center
\begin{tabular}{|c c|}
\hline
Porosity & $\varkappa = 0.7$ \\
Injection pressure & $p_I = 0.6$ MPa \\
Vent pressure & $p_0 = 0.1$ MPa\\
\hline
\end{tabular}
\caption{Physical properties of the preform.}\label{tab:properties}
\end{table}

\subsection{Convergence of the CV/FEM\label{sec:conv}}
 In this section we deal with the deterministic version of the model (\ref{two3}), i.e., when the hydraulic conductivity $K(x,y)$ is a given deterministic tensor.
 We apply Algorithm~\ref{alg:cvfem} to this model and examine its convergence looking at two quantities of interest:
 the filling time $\tau$ (it is not random in this experiment, of course) and volume of a void $v(t)$.

\begin{figure}[h!]
\centering
\subfloat[]{\includegraphics[width = .53\textwidth]{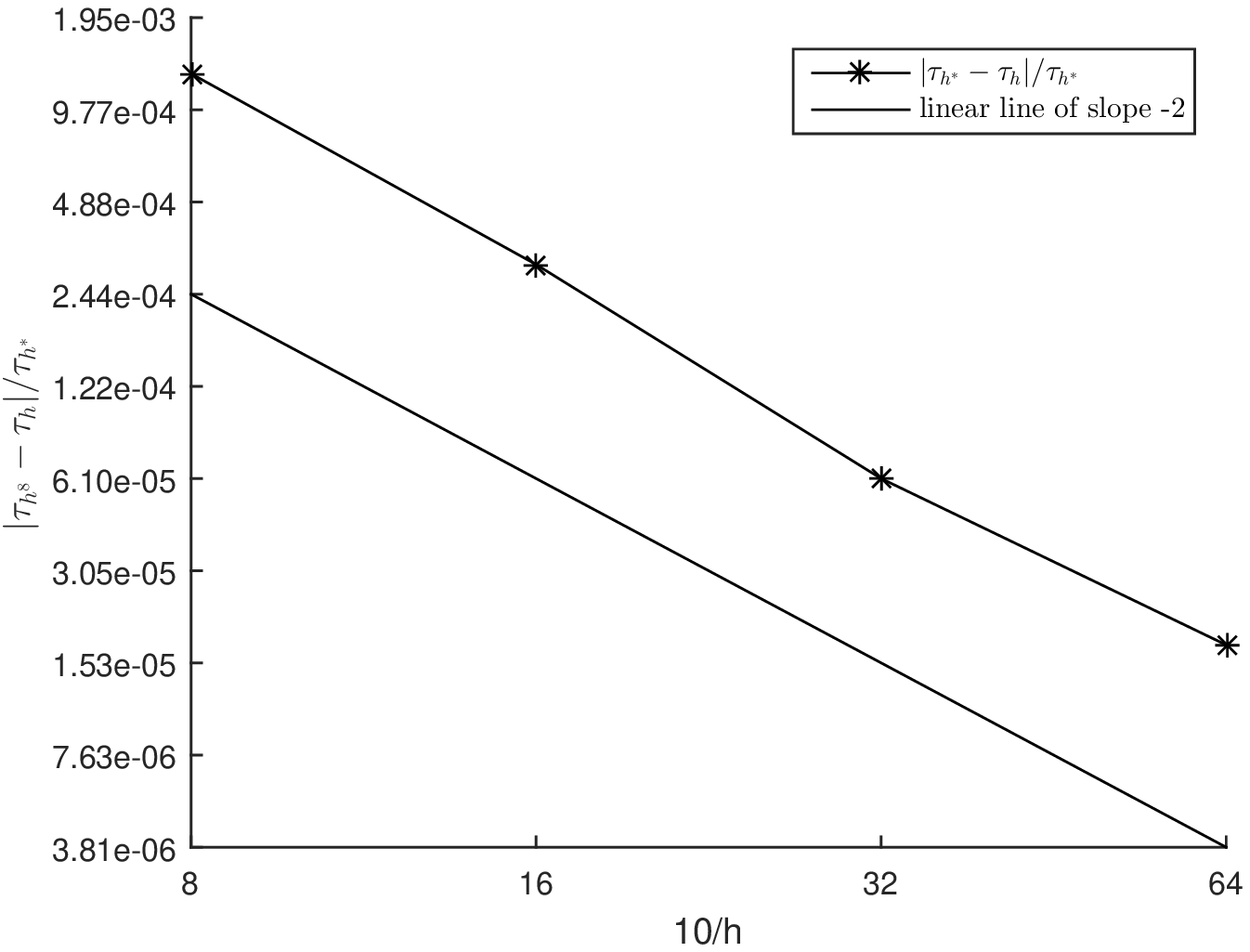}}
%\hspace{.5cm}
\subfloat[] {\includegraphics[width = .53\textwidth]{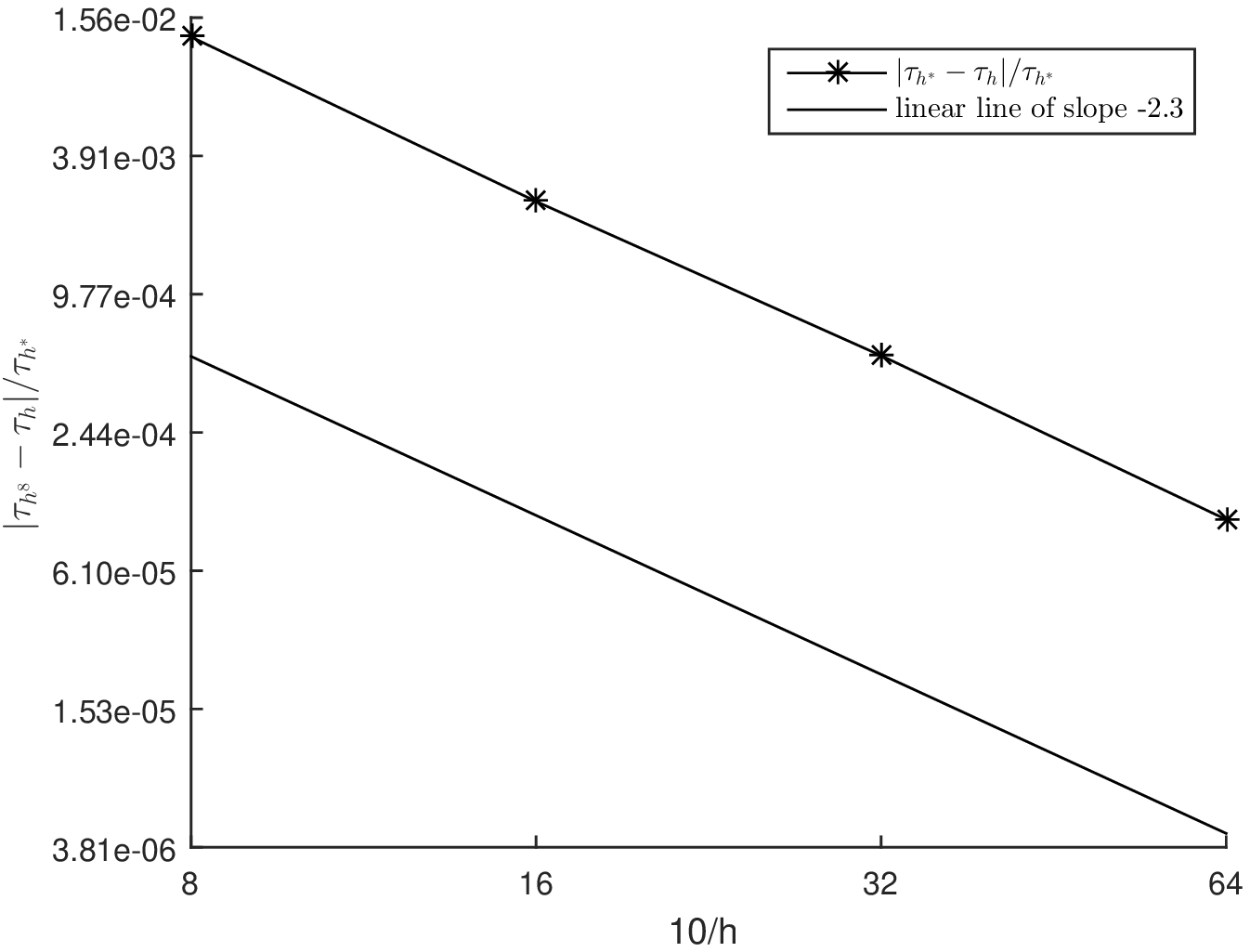}}
\hspace{.5cm}
\subfloat[] {
\includegraphics[width = .53\textwidth]{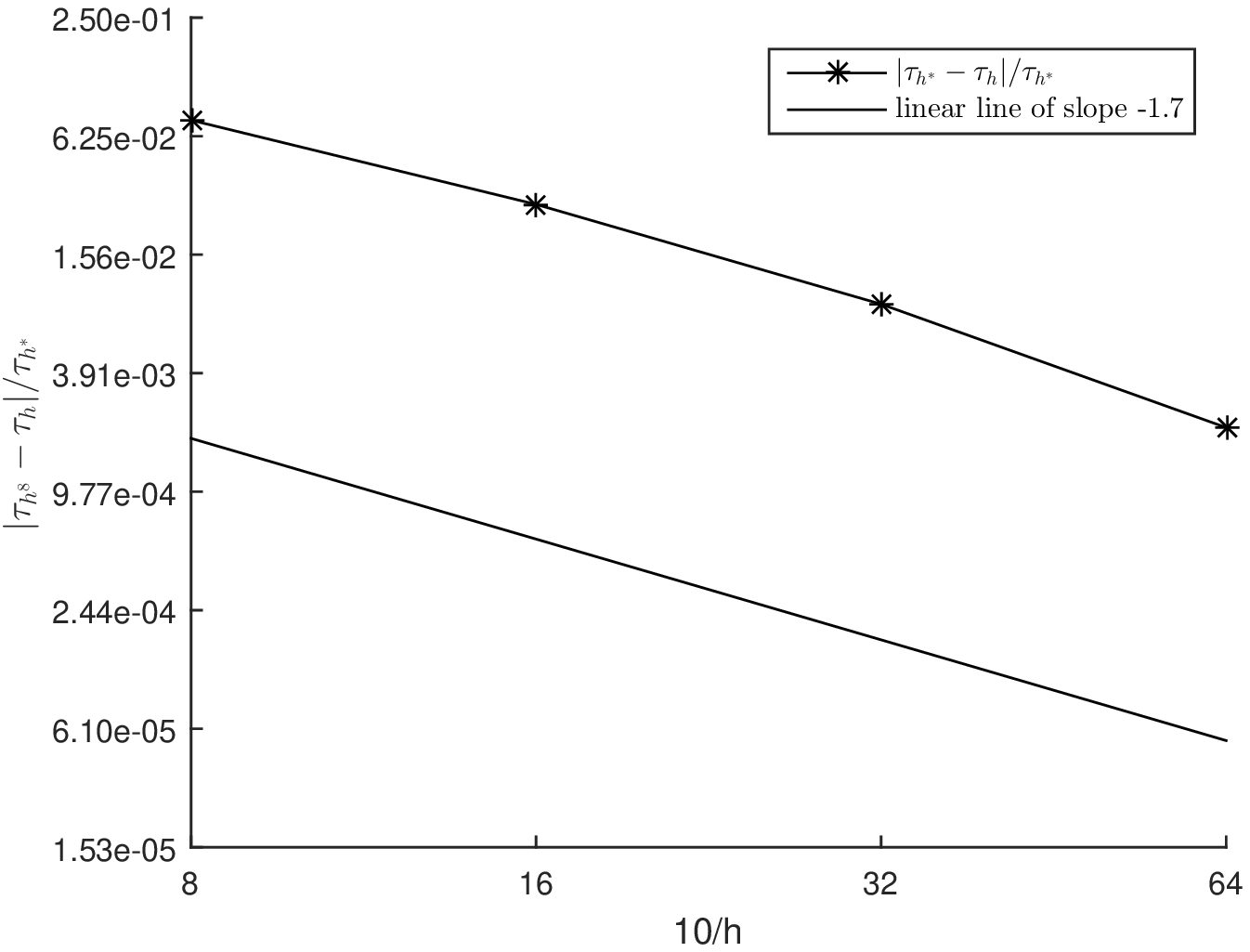}
}
\caption{The error of the filling time $|\tau_{h^*}-\tau_h|/\tau_{h^*}$ depending on the shapes of the flow front: (a) flat flow front,
(b) cup-shaped flow front, and (c) cap-shaped flow front.} \label{fig:convergence1}
\end{figure}

 To estimate the error of the simulated filling time $\tau_h$ obtained with the mesh size $h$, we use a reference solution $\tau_{h^*}$ obtained on a grid with the small mesh size $h^* = 10/128$ cm.
 Three different shapes of the flow front determined by different choices of the hydraulic conductivity $K(x,y)$ tensor are considered:
 (a) straight line parallel to y-axis ($K_{xx} = K_{yy} = 10^{-8} $ m$^2$/sec$\cdot$Pa);
 (b) cup-shaped front ($K_{xx}(x,y) = 10^{-9} +10^{-8} (1-\sin(\pi y/0.1))$ m$^2$/sec$\cdot$Pa, $K_{yy} = 10^{-10}  $ m$^2$/sec$\cdot$Pa);
  and (c) cap-shaped front  ($K_{xx}(x,y) = 10^{-9} +10^{-8} \sin(\pi y/0.1)$ m$^2$/sec$\cdot$Pa,
  $K_{yy} = 10^{-10}$ m$^2$/sec$\cdot$Pa). The convergence of the resin filling time $\tau_h$ with respect to the mesh resolution $h$ is shown in Fig.~\ref{fig:convergence1}. We observe that $|\tau_{h^*}-\tau_h|/\tau_{h^*}$ converges approximately quadratically in all three cases.
  At the same time, we see that the geometry of the flow front has an influence on the accuracy. The most accurate results are seen in the case of the flat flow front (Figure \ref{fig:convergence1} (a)) and the least accurate results are seen in the case of  the cap-shaped front (Figure \ref{fig:convergence1} (c)).

Void formation in RTM processes has received considerable interest due to its effect on the degradation of physical and mechanical properties of the composite.
According to \cite{JW78,HAA04}, even a void content of just $1\%$ can substantially affect mechanical properties of the material,
e.g., decrease of strength up to $30\%$ in bending, $9\%$ in torsional shear, $8\%$ in impact, etc.
Consequently, we view that it is important to look at convergence of the CV/FEM for the void content, despite not considering void
formation in further experiments here.

Let us look at accuracy of the CV/FEM with presence of a void.
In order to have a void in the solution of (\ref{two3}), we incorporate a low permeability patch ($1cm\leq x \leq 3cm$ and $4cm\leq y \leq 6cm$), where the permeability value is 100 times lower than that of the rest of the mold, $10^{-7}$  m$^2$/sec$\cdot$Pa (see also the corresponding discussion in Section~\ref{sec:2d}). In this setting only a single void can form.
We look at the void volume $\bar v(t)$ at the time $t^*$ when it is formed and at the time $t_{end}$ when the mold is fully saturated
except the void.

\begin{table}[h!]
\center
\begin{tabular}{|c| c c c|}
\hline
$10/h $ & $\bar{v}(t^*)\frac{p_0}{p_I} $ & $\bar{v}(t_{end})$ & $\bar{v}(t^*)$ \\ \hline

8 &2.04e-08 & 2.10e-08& 4.07e-08\\
16 & 6.84e-08& 7.09e-08& 1.37e-07\\
32 & 8.57e-08& 8.85e-08& 1.71e-07\\
64 & 9.23e-08& 9.54e-08& 1.86e-07\\
\hline
\end{tabular}
\caption{The volume of void (in $m^3$)
at $t^*$ and $t_{end}$ and also the lower bound for $\bar v (t_{end})$ as in (\ref{ieq1}).
 Here $p_I$ = 0.6 MPa and $p_0$ = 0.3 MPa.}\label{tab:upper_lower_bounds}
\end{table}

\begin{figure}[h!]
\centering
\subfloat[]{\includegraphics[width = .53\textwidth]{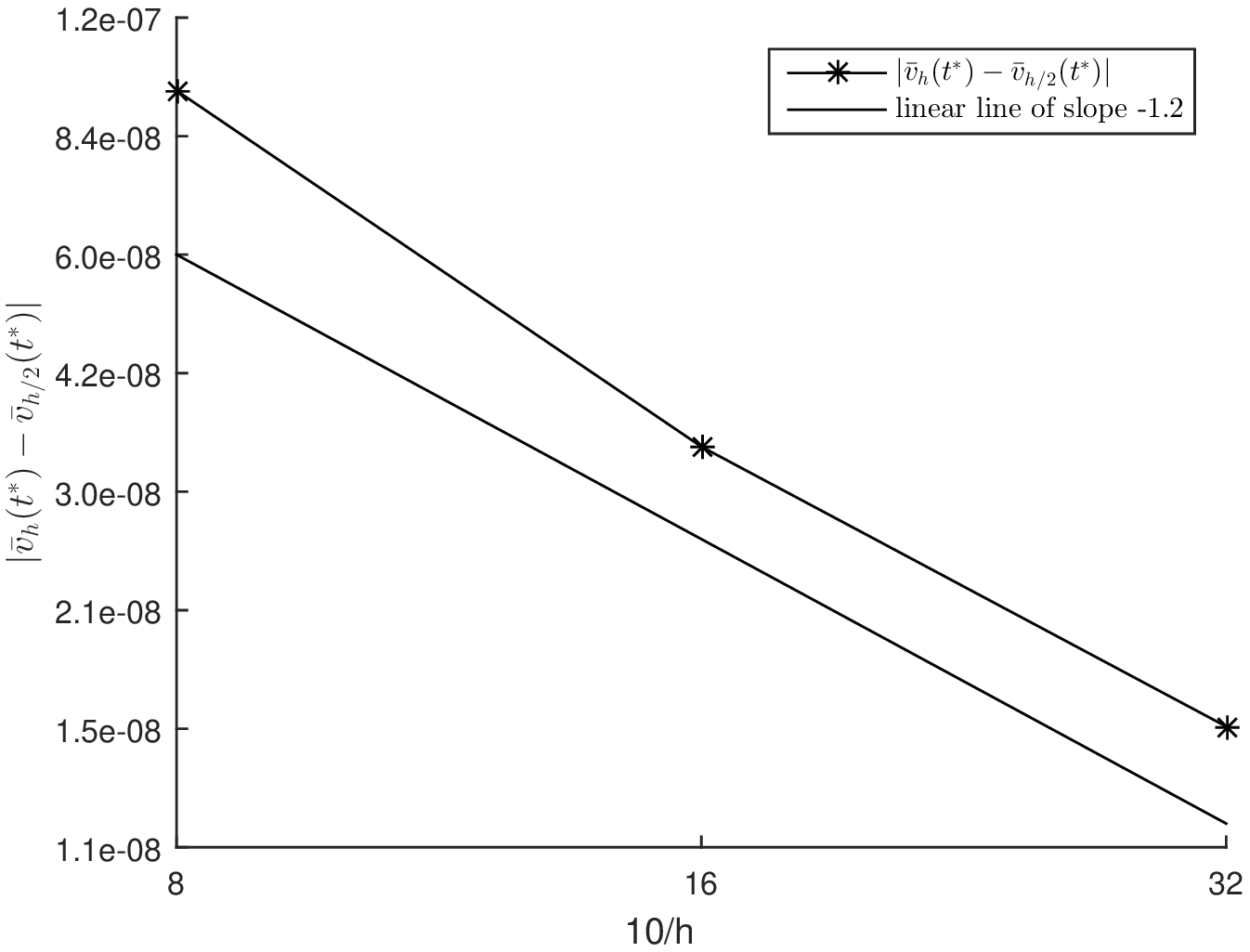}}
%\hspace{.1cm}
\subfloat[] {\includegraphics[width = .53\textwidth]{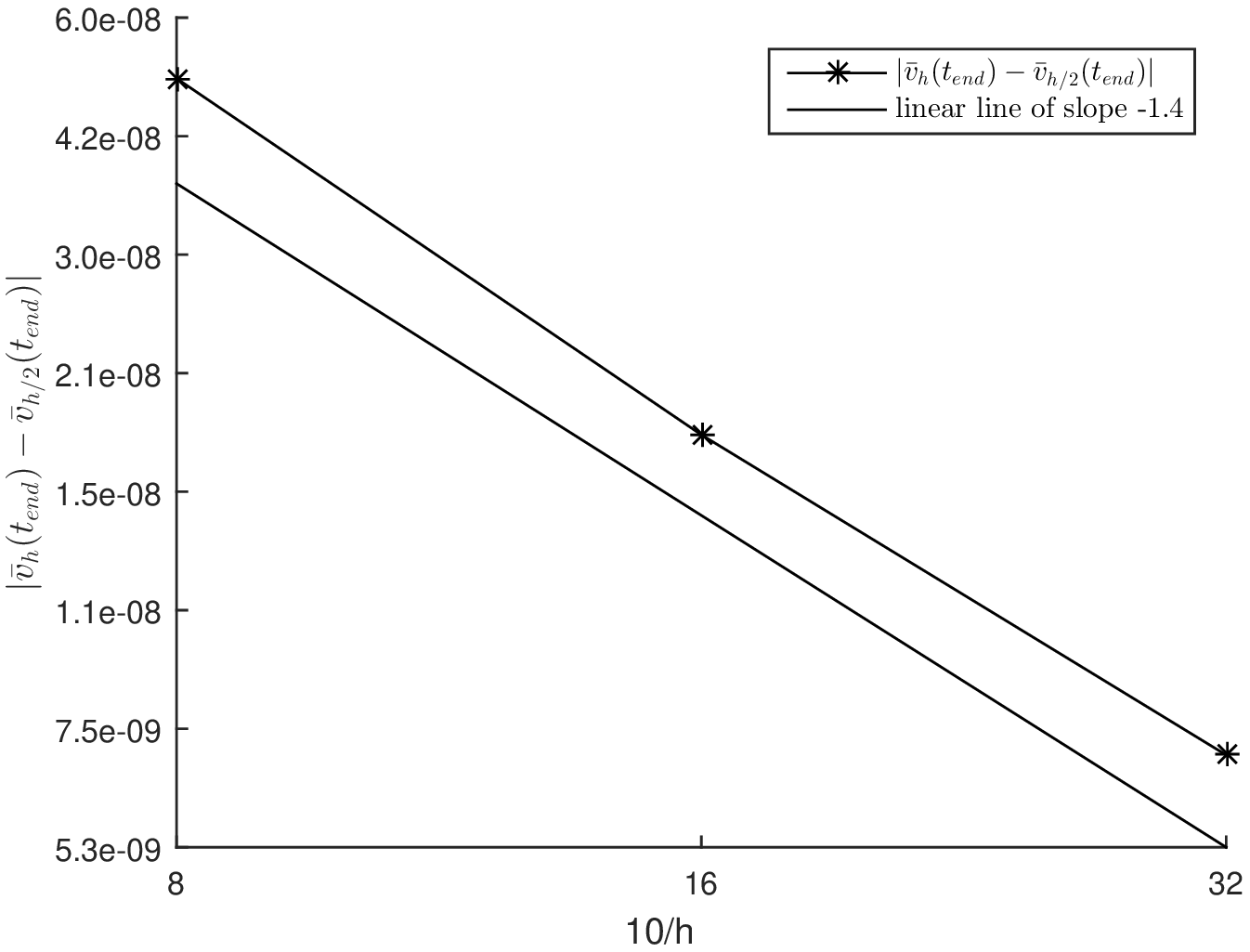}}
\caption{Void size difference between two consecutive discretization levels (a) at time $t^*$ when the void is formed for the first time and
 (b) at time $t_{\mathrm{end}}$ when the mold is completely filled except the void. } \label{fig:convergence2}
\end{figure}

Table~\ref{tab:upper_lower_bounds} shows that the inequality (\ref{ieq1}) holds in the discrete case.
The volumes of void at two times, $t^*$ and $t_{end}$, are both increasing as the mesh size $h$ becomes smaller.
Two plots in Fig.~\ref{fig:convergence2} show that the volume of void converges with approximately first order in $h$.

To conclude, we experimentally observed approximately
 2nd order convergence of the CV/FEM Algorithm~\ref{alg:cvfem} for the filling time
 and a lower order, approximately 1st order, convergence for the volume of void.

\subsection{Pseudo-1D flow}\label{sec:num1d}
  Now we consider the model (\ref{two3}), (\ref{two4}) with relatively high mean horizontal conductivity $\mu_{K_{xx}}$
  and low mean vertical conductivity $\mu_{K_{yy}}$ (i.e., $\mu_{K_{xx}} >> \mu_{K_{yy}}$), which results in limited movement of flow in the vertical direction. We also choose the horizontal correlation length $\lambda_x $
  to be much greater than the vertical correlation length $\lambda_y$ (i.e., $\lambda_x >> \lambda_y$). The parameters of the random conductivity tensor $K(x,y)$ used in this subsection are listed in Table \ref{tab:spar1}. We conduct 9 separate numerical experiments using the 3 different values of $\lambda_x$ for each of the three different Mat\'{e}rn  covariance functions with different smoothness $\nu$.

\begin{table}[h!]
\center
\begin{tabular}{|c r|c r|}
\hline
$\mu_{K_{xx}}$& 1e-8$\mathrm{m}^2/ \mathrm{sec}\cdot \mathrm{Pa}$ & $\mu_{\theta}$& 0 radian\\
$\sigma_{K_{xx}}$& 8e-9$\mathrm{m}^2/ \mathrm{sec}\cdot \mathrm{Pa}$&$\sigma_{\theta}$& 0.0356 radian\\
$\mu_{K_{yy}}$& 1e-9$\mathrm{m}^2/ \mathrm{sec}\cdot \mathrm{Pa}$&$\lambda_x$ & 0.01m, 0.03m, 0.05 m\\
$\sigma_{K_{yy}}$& 8e-10$\mathrm{m}^2/ \mathrm{sec}\cdot \mathrm{Pa}$&$\lambda_y$ & 0.001m\\
\hline
\end{tabular}
\caption{The parameters for the random hydraulic conductivity tensor $K(x,y)$ used in the pseudo-1D flow simulation.}\label{tab:spar1}
\end{table}

\begin{figure}
\centering
\includegraphics[width=0.45\textwidth]{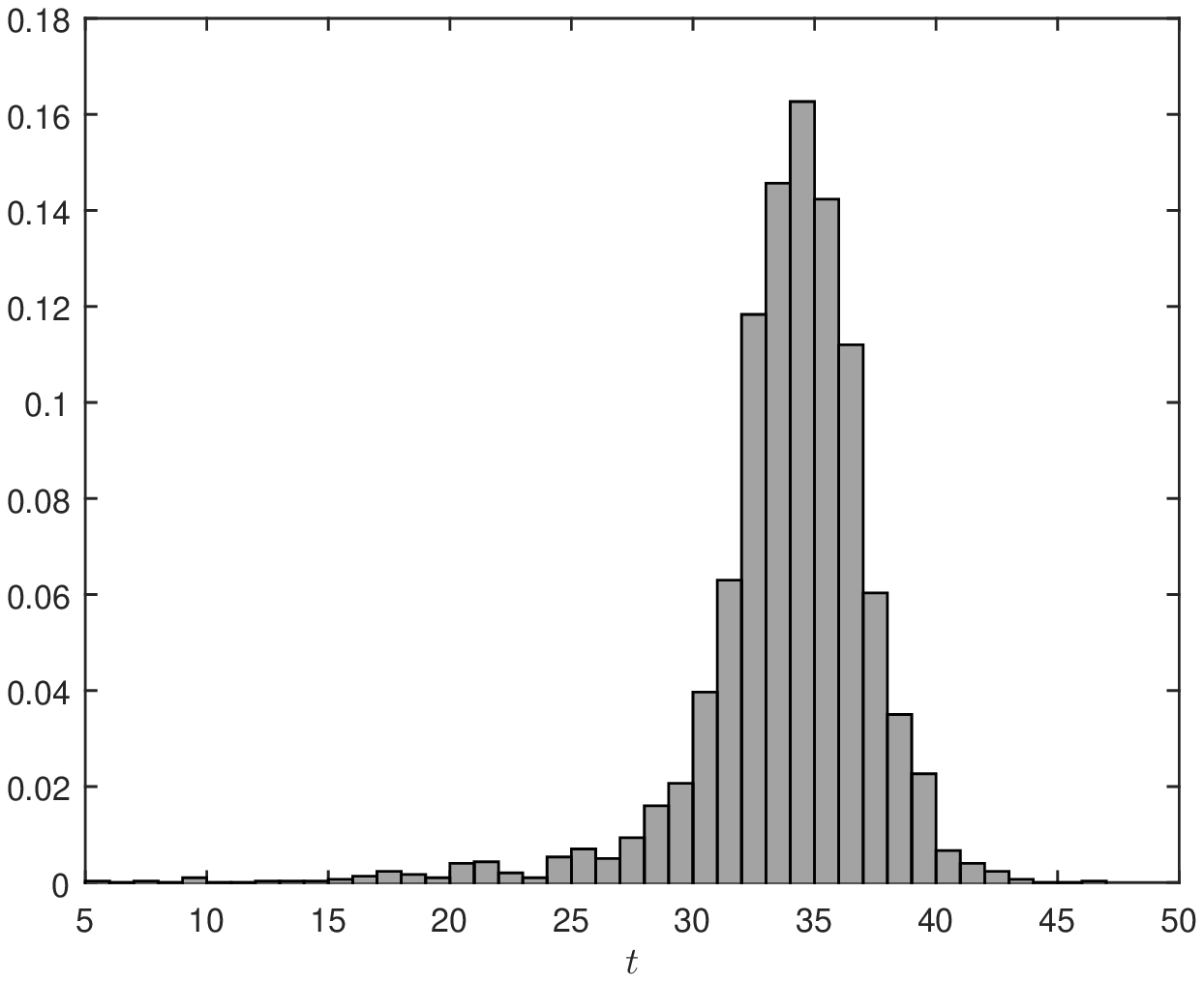}
\includegraphics[width=0.45\textwidth]{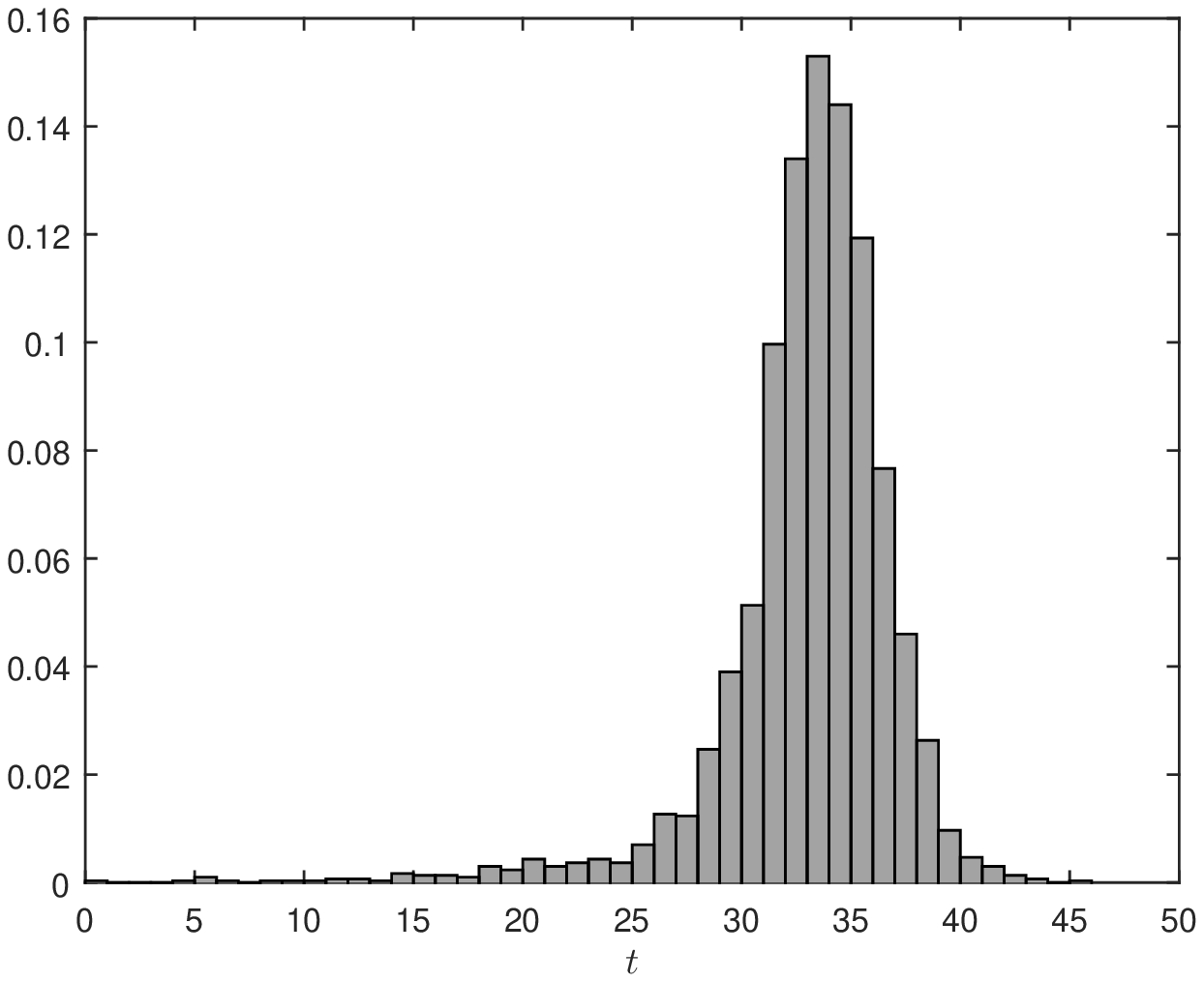}
\includegraphics[width=0.45\textwidth]{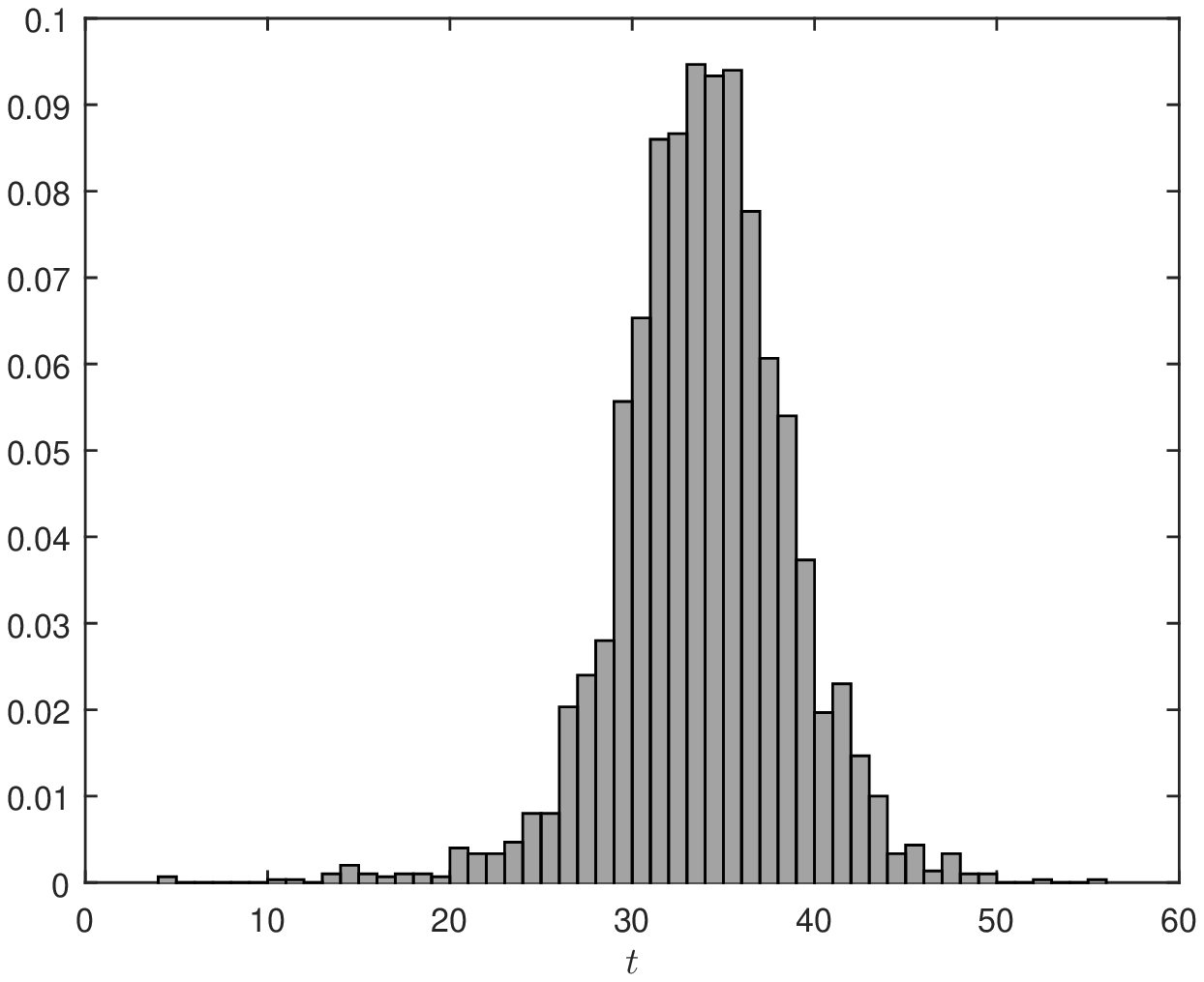}
\includegraphics[width=0.45\textwidth]{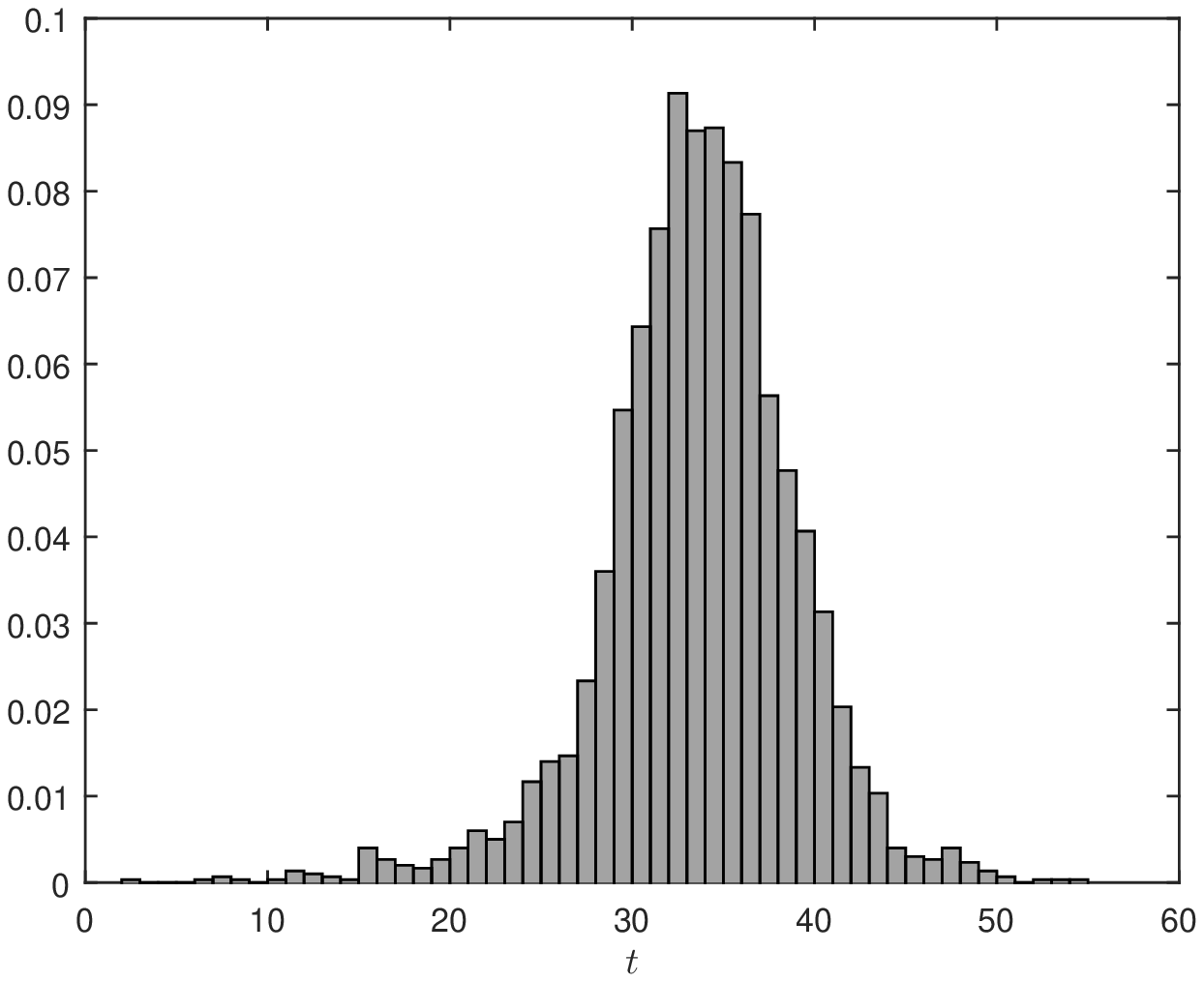}
\caption{ Sample densities for $\tau$ (in sec) in the case of pseudo-1D flow. The left top figure: $\nu=1/2$, $\lambda=0.01$;
the right top figure: $\nu=5/2$, $\lambda=0.01$; left bottom figure: $\nu=1/2$, $\lambda=0.05$;
right bottom figure: $\nu=5/2$, $\lambda=0.05$.}
\label{fig:den}
\end{figure}

The results of the experiments are presented in Tables~\ref{tab:mean_tau1} and~\ref{tab:var_tau1}.  We observe that the mean filling time $E\tau$ shows almost no dependence on either the horizontal correlation length $\lambda_x$ or on smoothness $\nu$ of the random hydraulic conductivity field $K(x,y)$, while the variance of the filling time increases with increase of $\lambda_x$ and $\nu$. This is consistent with the 1D flow properties studied in Section~\ref{sec:1d} and one can conclude that when $\mu_{K_{xx}} >> \mu_{K_{yy}}$, a good qualitative prediction about the filling time $\tau$ can be made using the analytically solvable one-dimensional problem (\ref{one1}). Note that the filling time according to the design here and in the corresponding one-dimensional case (see Section~\ref{sec:1d}) coincide as expected but it is not so
for $E\tau$ (see Remark~\ref{rem:dif} below). We also plot typical sample densities for $\tau$ (see Fig.~\ref{fig:den}). We observe that the probability of filling time $\tau$ being in a range close to the designed time (2.8 sec here) is very low.

\begin{table}[h!]
\center
\begin{tabular}{|c|c|c|c|}
\hline
&  \multicolumn{3}{|c|}{$E\tau$}  \\
\hline
$\lambda_x$ & $\nu = 1/2$ &  $\nu = 3/2$ &  $\nu = 5/2$  \\
\hline
0.01 & 33.9$\pm$0.1& 33.3$\pm$0.1&33.1$\pm$0.1 \\
\hline
0.03 & 33.8$\pm$0.2& 33.4$\pm$0.2&33.4$\pm$0.2\\
\hline
0.05 & 33.9$\pm$0.2& 33.6$\pm$0.2 &33.7$\pm$0.2\\
\hline
\end{tabular}
\caption{Mean filling time $E\tau$ (in sec) and its 95$\%$ confidence interval for the pseudo-1D flow.
The filling time according to the design is equal to 2.8 sec.}\label{tab:mean_tau1}
\end{table}

\begin{remark}
We note that though the filling times according to the design in the considered two-dimensional case here and in the one-dimensional case are both equal to $2.8$ sec, there is a considerable difference between $E\tau$ in the two cases: in the pseudo-1D flow $E\tau \approx 34$ sec, while in the true 1-D case (see Section~\ref{sec:1d}) $E\tau \approx 4.6$ sec. The reason for this disparity of the mean filling times in 1D and 2D cases can be illustrated in the following way. Let us imagine the extreme case of the pseudo-1D flow so that $K_{yy}$=0 and $\theta=0$. Suppose in our domain discretization we have $l$ horizonal strips via which resin is propagating. Each strip has its own time-to-fill $\tau_i$ on each realization of the random field $K(x)$. Then for $\tau=\max_{1 \leq i \leq l} \tau_i$ we have $E \tau \geq E \tau_i$.  It is an interesting probabilistic mini-problem to study the relationship between $\tau$ and $\tau_i$, in particular between $E\tau$ and $E\tau_i$,
  as well as the limiting case $l\rightarrow \infty$, but we do not consider it here. We emphasize that in the deterministic case with $K_{xx} >> K_{yy}$ the one-dimensional model gives good predictions for the two-dimensional model and, being considerably simpler, it is often used in practice for this purpose. However, here we highlight that in the stochastic case there are considerable quantitative differences between statistical characteristics of the pseudo-1D flow and the 1D flow despite the fact that qualitatively they are similar. This observation is important from the practical point of view. Further, in \cite{LPT05,XT06} expansions of moments of the interface dynamics were derived using a dynamical mapping of the Cartesian coordinate system onto a coordinate system associated with the moving front. Making use of the ideas of \cite{LPT05} to obtain expansions for variance of filling times is an interesting problem (even if we assume that discontinuity of the front due to possible void formation can be neglected) for future research.
\label{rem:dif}
\end{remark}

\begin{table}[h!]
\center
\begin{tabular}{|c|c|c|c|}
\hline
&  \multicolumn{3}{|c|}{$\mathrm{Var}\tau$}  \\
\hline
 $\lambda_x$ & $\nu = 1/2$  &  $\nu = 3/2$ & $\nu = 5/2$\\
 \hline
 0.01 & 12.9$\pm$1.5& 13.5$\pm$1.8& 15.3$\pm$1.9\\
\hline
0.03 & 18.8$\pm$1.5& 20.0$\pm$1.6& 20.4$\pm$1.5\\
\hline
0.05 & 22.7$\pm$1.7& 26.2$\pm$2.0& 28.4$\pm$2.2\\
\hline
\end{tabular}
\caption{Variance of the filling time $\mathrm{Var}\tau$ (in sec$^2$) and its 95$\%$ confidence interval for the pseudo-1D flow.} \label{tab:var_tau1}
\end{table}

\subsection{Two-dimensional isotropic flow}\label{sec:num2d}
The second stochastic numerical experiment uses the following parameters of the random hydraulic conductivity: $\mu_{K_{xx}} =\mu_{K_{yy}} = 1e$-$8$$\mathrm{m}^2/ \mathrm{sec}\cdot \mathrm{Pa}$, $\sigma_{K_{xx}} = \sigma_{K_{yy}} = 8e$-$9$$\mathrm{m}^2/ \mathrm{sec}\cdot \mathrm{Pa}$, $\mu_\theta = 0$, $\sigma_\theta = 0.0356$ radians, and  $\lambda_x = \lambda_y = $ \{0.01m, 0.03m, 0.05m\}. The focus of this experiment is again to investigate the impact of the correlation length and smoothness of
the random field $K(x,y)$ on the filling time.

\begin{table}[h!]
\center
\begin{tabular}{|c|c|c|c|}
\hline
&  \multicolumn{3}{|c|}{$E\tau$}  \\
\hline
$\lambda_x = \lambda_y$ & $\nu = 1/2$ &  $\nu = 3/2$ &  $\nu = 5/2$  \\
\hline
0.01 & 37.8$\pm$0.2& 39.4$\pm$0.2&39.8$\pm$0.2\\
\hline
0.03 & 42.1$\pm$0.5& 43.6$\pm$0.5&43.3$\pm$0.5\\
\hline
0.05 & 43.3$\pm$0.6& 45.2$\pm$0.7 &45.4$\pm$0.7\\
\hline
\end{tabular}
\caption{Mean filling time $E\tau$ (in sec) and its 95$\%$ confidence interval for the 2D isotropic flow.
The filling time according to the design is equal to 2.8 sec.}\label{tab:mean_tau2}
\end{table}

\begin{table}[h!]
\center
\begin{tabular}{|c|c|c|c|}
\hline
&  \multicolumn{3}{|c|}{$\mathrm{Var}\tau$}  \\
\hline
 $\lambda_x = \lambda_y$ & $\nu = 1/2$  &  $\nu = 3/2$ & $\nu = 5/2$\\
 \hline
 0.01 &32.4$\pm$2.6& 35.1$\pm$2.5& 35.4$\pm$2.4\\
\hline
0.03 & 170$\pm$12& 216$\pm$17& 219$\pm$20\\
\hline
0.05 & 300$\pm$23& 432$\pm$41& 424$\pm$41\\
\hline
\end{tabular}
\caption{Variance of the filling time $\mathrm{Var}\tau$ (in sec$^2$) and its 95$\%$ confidence interval for the 2D  mean isotropic flow.}
\label{tab:var_tau2}
\end{table}

Unlike the pseudo-1D results presented in Section~\ref{sec:num1d}, where the mean filling time $E\tau$ is independent of the correlation length and smoothness,
here $E\tau$  increases with growth of the correlation length as shown in Table~\ref{tab:mean_tau2} and slightly grows with increase of smoothness.
In other words, 2D flow moves slower on a more homogeneous porous medium with less spatial variability than in the pseudo-1D case.
We also observe a fast increase of variance $\mathrm{Var}\tau$ with an increasing spatial correlation length in Table \ref{tab:var_tau2}.
%
%We note that we see here formation of voids despite the design being very
%simple. We expect that random deviations from a more complex design have
%even more drastic effects.
%

\subsection{Two-dimensional flow with anisotropic mean}\label{sec:num2d2}
The third stochastic numerical experiment uses the following parameters of the random hydraulic conductivity:
 $\mu_{K_{xx}} = 1e$-$8$$\mathrm{m}^2/ \mathrm{sec}\cdot \mathrm{Pa}$, $\mu_{K_{yy}} = 2e$-$8$$\mathrm{m}^2/ \mathrm{sec}\cdot \mathrm{Pa}$,
 $\sigma_{K_{xx}} = \sigma_{K_{yy}} = 8e$-$9$$\mathrm{m}^2/ \mathrm{sec}\cdot \mathrm{Pa}$, $\mu_\theta = 0$, $\sigma_\theta = 0.0356$ radian, and
 $\lambda_x = \lambda_y = $ \{0.01m, 0.03m, 0.05m\}.
 In other words, here (in comparison with Section~\ref{sec:num2d})
 we consider a model in which  random hydraulic conductivity has an anisotropic mean.
 Recall (see Section~\ref{sec:2d}) that we proposed that the means of $K_{xx}(x,y),$ $K_{yy}(x,y)$ and $\theta (x,y)$ correspond
to the hydraulic conductivity intended by the design of a composite part and that
in most cases of practical interest permeability $k(x,y)$ (and hence the
hydraulic conductivity $K(x,y)$) is anisotropic by design.
Note that the mean of permeability in the $x$-direction ($K_{xx}$) is smaller than that in the $y$-direction ($K_{yy}$) and that the 1D model
from Section~\ref{sec:1d} does not apply here (see the discussion in Section~\ref{sec:num1d}).

The results of the numerical experiments are presented in Tables~\ref{tab:mean_tau3} and~\ref{tab:var_tau3}.
We observe that the mean filling time $E\tau$ of the 2D mean anisotropic flow increases as the correlation length increases and also that
an increasing spatial correlation length leads to an  increase of $\mathrm{Var}\tau$.

\begin{table}[h!]
\center
\begin{tabular}{|c|c|c|c|}
\hline
&  \multicolumn{3}{|c|}{$E\tau$}  \\
\hline
$\lambda_x = \lambda_y$ & $\nu = 1/2$ &  $\nu = 3/2$ &  $\nu = 5/2$  \\
\hline
0.01 & 35.4$\pm$0.2& 36.8$\pm$0.2&37.0$\pm$0.2\\
\hline
0.03 & 39.3$\pm$0.4& 40.1$\pm$0.5&40.3$\pm$0.5\\
\hline
0.05 & 40.1$\pm$0.6& 42.1$\pm$0.7 &42.1$\pm$0.7\\
\hline
\end{tabular}
\caption{Mean filling time $E\tau$ (in sec) and its 95$\%$ confidence interval for the 2D mean anisotropic flow.
The filling time according to the design is equal to 2.64 sec.}\label{tab:mean_tau3}
\end{table}

We note that in all three stochastic cases considered in Sections~\ref{sec:num1d}-\ref{sec:num2d2} the mean filling time is considerably higher that
the filling time according to the design, which further emphasises the importance of stochastic modeling of RTM.

\begin{table}[h!]
\center
\begin{tabular}{|c|c|c|c|}
\hline
&  \multicolumn{3}{|c|}{$\mathrm{Var}\tau$}  \\
\hline
 $\lambda_x = \lambda_y$ & $\nu = 1/2$  &  $\nu = 3/2$ & $\nu = 5/2$\\
 \hline
 0.01 &24.9$\pm$2.0& 26.1$\pm$1.7& 27.9$\pm$2.0\\
\hline
0.03 & 135$\pm$9& 185$\pm$26&178$\pm$13\\
\hline
0.05 & 257$\pm$23& 355$\pm$33&353$\pm$27\\
\hline
\end{tabular}
\caption{Variance of the filling time $\mathrm{Var}\tau$ (in sec$^2$)
and its 95$\%$ confidence interval for the 2D mean anisotropic flow.} \label{tab:var_tau3}
\end{table}

%\subsection{CVFEM convergence issue with voids}
%\update{1.3}{We have shown the effects of the correlation length and the smoothness of random fields on the mean fill time $E\tau$ and the variance of the fill time $\mathrm{Var}\tau$.
%In this section, we study the convergence of the CV/FEM solution with random hydraulic conductivity. To begin with, we first provide the convergence results of the CV/FEM in two deterministic cases. }
%\update{1.3}{The first case is that the hydraulic conductivity is isotropic with $K_{xx} = K_{yy} = 1e-8 \mathrm{m}^2$  and $\theta = 0$. The mesh size $h$ is the length of hypotenuse of the triangular element. To estimate the errors, we approximate the exact solution $\tau$ by a reference solution $\tau_{1/h^*}$ on a grid with mesh size $h^* = 10/128$cm. We observe that $|\tau_{h^*} - \tau_{h}	|$ converges quadratically in $h$ as shown in Figure \ref{fig:conv_result_det} (a).}
%
%\update{1.3}{The second deterministic example contains a variation in the conductivity of the preform, which result in the void formation.  The preform has a square panel (5cm $\leq x \leq$ 10cm and 2.5cm$\leq y \leq$ 7.5cm), where the conductivity has been assigned to be $1e-10\mathrm{m}^2$, which is 100 times lower than the conductivity of the remainder of the computational domain. In Figure \ref{fig:conv_result_det} (b), we observe $\mathcal{O}({h^{3/2}})$ convergence for $|\tau_{h^*} - \tau_{h}	|$.
%}

\begin{remark}
The MATLAB codes for the CV/FEM used in our experiments are available at \allowbreak\textit{https://github.com/parkmh/MATCVFEM.}
\end{remark}
\section{Discussion and summary\label{sec:end}}
In this work we considered stochastic one-dimensional and two-dimensional moving-boundary problems suitable for modeling RTM processes.
The PDE formulation of the main two-dimensional model which takes into account compressible air entrapment in voids behind the main front
is somewhat novel. The one-dimensional problem has an analytical solution while the two-dimensional model requires a numerical method for
computing quantities of interest. Following the common practice in the RTM community \cite{AS10}, we use the control volume-finite element method (CV/FEM)
to simulate the two-dimensional problem. We test accuracy of the CV/FEM algorithm for three particular cases of deterministic hydraulic conductivity,
and we experimentally observed approximately its 2nd order convergence for the filling time and a lower order, approximately 1st order,
convergence for the volume of void. We note that these tests were done in the case of infinitely smooth (in the case of the filling time)
or piece-wise constant (in the case of the volume of void) hydraulic conductivity while less regular random field are used to model
hydraulic conductivity. For future study, it is of interest to look at dependence of behavior of the CV/FEM algorithm on the smoothness of
hydraulic conductivity and on various observables.

We studied properties of a stochastic one-dimensional moving-boundary problem with
the hydraulic conductivity being modelled as a stationary log-normal random field.
In particular, we observe that the mean filling time does not depend on correlation length or on smoothness of hydraulic conductivity,
while the filling time variance (as a measure of its variability) does depend on both the correlation length and smoothness. A similar conclusion is made about the two-dimensional model's behavior when random permeability in the direction from inlet to outlet is much bigger than in the perpendicular direction (the case of pseudo-1D flow).
However, we discovered that in other cases of random permeability the mean filling time does depend on correlation length and on smoothness.
The important consequence of these conclusions is the observed (often high) sensitivity of the mean and variance of the filling time to changes in correlation length of the permeability as well as in its smoothness. This highlights the importance of conducting laboratory experiments from which covariance of the permeability field can be reconstructed.
Further, sensitivity of filling time to smoothness of permeability serves as a warning that stochastic modeling of permeability via homogenization procedures needs to be done with a very careful choice of scales.

Among the main objectives of this paper was to attract attention of the UQ community to challenges posed by stochastic moving-boundary problems,
which are highly relevant to modern technological processes in production of composite materials as well as to other porous media problems.
The challenges include (i) establishing existence and uniqueness results for two and three dimensional moving-boundary problems of the type
considered in this work;
(ii) numerical analysis for the corresponding CV/FEM algorithm; (iii) development of faster sampling techniques, e.g. using a multi-level Monte Carlo
approach \cite{giles,muller1,muller2} and/or polynomial chaos expansions \cite{pce2,pce3,pce4};
(iv) computational experiments with complex geometry using outcomes of (iii);
(v) comparing the CV/FEM with other numerical approaches to moving boundary problems, e.g. level sets methods \cite{levelset};
(vi) design of laboratory experiments and collection and analysis of the corresponding data to recover characteristics
of random permeability and to find a stochastic model of the conductivity consistent with experimental data
(for recent research in this direction see e.g. \cite{AS10,MAL14,Matveev}); and (vii) considering a two-phase model involving both incompressible
(resin) and compressible (air) phases and comparing its properties with the ones for the reduced model studied here.

\section*{Acknowledgements}

This work was partially supported by the EPSRC grant EP/K031430/1. The
authors are very grateful to Matthew Hubbard, Arthur Jones, Andy Long, Alex Skordos, and Kris van der Zee for useful discussions.
We also express special thanks to Mikhail Matveev who read drafts of this paper and gave insightful feedbacks from
the engineering prospective.

\end{document}